\documentclass[superscriptaddress, prd, aps,amsmath,amssymb,showpacs,showkeys, onecolumn]{revtex4}
\usepackage[dvips]{graphicx,color}
\usepackage{times}
\usepackage{xcolor}
\usepackage{caption}
\usepackage[%
  colorlinks=true,
  urlcolor=blue,
  linkcolor=red,
  citecolor=blue
]{hyperref}

\begin{document}

\title{Optical, Dynamic and Topological Characteristics of Deformed Schwarzschild Black Holes
}

\author{Dhruba Jyoti Gogoi\footnote{Corresponding author}}
\email[Email: ]{moloydhruba@yahoo.in}
\affiliation{Department of Physics, Moran College, Moranhat, Charaideo 785670, Assam, India.}
\affiliation{Research Center of Astrophysics and Cosmology,
Khazar University, 41 Mehseti Street, AZ1096 Baku, Azerbaijan.}

\author{Jyatsnasree Bora}
\email[Email: ]{jyatnasree.borah@gmail.com}
\affiliation{Pacif Institute of Cosmology and Selfology (PICS), Sagara, Sambalpur 768224, Odisha, India.}

\author{Filip Studnička}
\email[Email: ]{filip.studnicka@uhk.cz}
\affiliation{Department   of   Physics, Faculty of Science,  University   of Hradec Kr$\acute{a}$lov$\acute{e}$, Rokitansk$\acute{e}$ho 62, 500 03 Hradec Kr$\acute{a}$lov$\acute{e}$, Czechia.}

\author{H. Hassanabadi}
\email[Email: ]{hha1349@gmail.com}
\affiliation{Department of Physics, Faculty of Science, University of Hradec Kr$\acute{a}$lov$\acute{e}$, Rokitansk$\acute{e}$ho 62, 500 03 Hradec Kr$\acute{a}$lov$\acute{e}$, Czechia.}
\affiliation{Departamento de F\'{\i}sica Te\'orica, At\'omica y Optica and Laboratory for Disruptive \\ Interdisciplinary Science (LaDIS), Universidad de Valladolid, 47011 Valladolid, Spain}

\begin{abstract}

We investigate the temperature, photon and shadow radii, quasinormal modes (QNMs), {time domain profiles}, greybody factors, emission rates and topological characteristics of deformed black holes, focusing on the effects of the deformation parameter $ \alpha $ and control parameter $ \beta $. Increasing $ \alpha $ enhances the oscillation frequency and damping rate of gravitational waves, while $ \beta $ shows non-linear behaviour. Electromagnetic perturbations exhibit similar trends, though with lower frequencies and damping rates. Greybody factors are mainly influenced by multipole moment $l$ and $ \alpha $, with $ \beta $ having a more subtle effect. These findings provide insights into black hole dynamics, mergers, and gravitational wave emissions.

\end{abstract}

\maketitle
\section{Introduction}\label{Sec1}

The study of black holes has captivated scientists for decades due to their extreme gravitational properties, but their importance extends beyond gravity. Black holes are now recognized as crucial objects for understanding thermodynamic principles in the most extreme conditions of the universe. Black hole thermodynamics, a field that merges the principles of classical thermodynamics with general relativity, quantum mechanics, and statistical mechanics, has emerged as a fascinating and vital area of research. It not only deepens our understanding of black hole behaviour but also serves as a potential gateway to formulating a comprehensive theory of quantum gravity.

Black hole thermodynamics began with a groundbreaking insight by J. D. Bekenstein, who, in the early 1970s, proposed that black holes should possess entropy \cite{Bekenstein:1972tm}. His work established a direct relationship between the area of a black hole’s event horizon and its entropy, indicating that the surface area of the horizon could be a measure of the disorder or information content of a black hole \cite{Bekenstein:1972tm}. This idea transformed the way scientists viewed black holes, not merely as objects that consume matter and energy, but as thermodynamic entities that obey laws analogous to those of classical thermodynamics.

Building upon Bekenstein's work, Stephen Hawking made a revolutionary discovery in 1974 when he demonstrated that black holes can emit radiation, now famously known as Hawking radiation. This radiation arises from quantum mechanical effects near the event horizon, where particle-antiparticle pairs are created, and one of the particles escapes while the other falls into the black hole. Over time, this process causes the black hole to lose mass and eventually evaporate \cite{Hawking:1975vcx}. Hawking's discovery challenged the traditional notion of black holes as objects from which nothing can escape and introduced a connection between black hole thermodynamics and quantum mechanics. This discovery also implied that black holes possess a finite temperature, related to their mass and surface area, adding to the richness of their thermodynamic properties \cite{Wald:1999vt}.

The relationship between the area of a black hole’s event horizon and its entropy, formulated by Bekenstein and Hawking, is now considered a cornerstone of black hole thermodynamics. Their collaboration helped define the four laws of black hole thermodynamics, which closely parallel the classical laws of thermodynamics. These laws govern the behaviour of black hole systems in terms of temperature, entropy, and energy conservation \cite{Bekenstein:1973ur,Wald:1999vt,Bardeen:1973gs,Georgescu:2021dyo}. Specifically, the second law of black hole thermodynamics, which states that the total entropy of a black hole cannot decrease, is analogous to the second law of thermodynamics, which dictates that the total entropy of a closed system never decreases.

As the field progressed through the 1980s and 1990s, black hole thermodynamics became more intricately connected with quantum mechanics. One of the most significant developments during this period was the introduction of the holographic principle, which postulates that all the information contained within a volume of space can be represented on the boundary of that space. This idea suggests that the physics inside a black hole, or even a more general region of spacetime, can be understood in terms of information encoded on its boundary, hinting at deeper connections between gravity and quantum field theory \cite{Wald:1999vt}. This concept found concrete application in the AdS/CFT correspondence, a powerful tool in theoretical physics that relates a gravitational theory in {Anti-de Sitter} (AdS) space to a conformal field theory (CFT) on its boundary. Such correspondences have proven invaluable in studying black hole thermodynamics and quantum gravity.

In recent years, {researchers have} continued to explore black hole thermodynamics across a variety of theoretical frameworks, including modified theories of gravity such as string theory and loop quantum gravity \cite{Ashtekar:2004eh,Dvali:2011aa}. These modified theories often lead to generalizations or deformations of the traditional thermodynamic laws that apply to black holes. For example, when incorporating the generalized uncertainty principle, a modification of the Heisenberg uncertainty principle to account for quantum gravitational effects, the entropy and temperature of black holes are modified, leading to potentially observable consequences \cite{Amelino-Camelia:2000stu,Ali:2009zq,Yang:2023nnk}. These generalized uncertainty principles are particularly relevant in the context of small black holes or black holes that form in high-energy regimes, such as those potentially produced in particle accelerators.

Furthermore, external factors such as electromagnetic fields or the presence of a cosmological constant can significantly alter the thermodynamic behaviour of black holes. For instance, the presence of a cosmological constant, which leads to Anti-de Sitter or de Sitter space solutions, introduces new thermodynamic properties and phase transitions in black holes. Such phase transitions, which are akin to those found in classical thermodynamics (such as the transition between solid, liquid, and gas phases), are observed in various black hole solutions, including charged, rotating, and higher-dimensional black holes \cite{Kubiznak:2012wp,Hendi:2012um,Teitelboim:1985dp,Gunasekaran:2012dq,Cvetic:1999ne,Chamblin:1999tk}. In these systems, changes in parameters like charge, angular momentum, or cosmological constant can lead to critical phenomena and phase transitions, which have been a topic of intense study in recent years.

For example, research on rotating Kerr-AdS black holes using holographic techniques has provided {insight} into their phase transitions and thermodynamic properties \cite{Gong:2023ywu}. Similarly, accelerating black holes and those in lower-dimensional spacetimes offer simplified contexts for studying thermodynamic stability and transitions, contributing to the broader understanding of black hole microstructures \cite{Sokoliuk:2023pby, Singh:2023hit, Blagojevic:2023kqj, Xiao:2023lap, Wu:2022whe}. Such studies are crucial for understanding the nature of black hole entropy and the information paradox, one of the most significant unresolved issues in theoretical physics.
{
Phase transitions and critical phenomena have been extensively investigated in black hole thermodynamics \cite{Gunasekaran:2012dq,Cvetic:1999ne,Chamblin:1999tk}. Various types of black holes in different modified gravity theories have also been found to exhibit a wide range of phase transitions \cite{Wei:2012ui,Altamirano:2014tva}.

Through holographic techniques, rotating Kerr-AdS black holes have been analyzed for their phase transitions and thermodynamic properties \cite{Gong:2023ywu}, while three-dimensional accelerating black holes provide a more simplified context for such studies \cite{Tian:2023ine}. The authors of Ref. \cite{Sokoliuk:2023pby} examined AdS black holes, focusing on their microstructures and stability, contributing to insights related to the AdS/CFT correspondence. In Ref. \cite{Singh:2023hit}, thermodynamic curvature transitions across different ensembles in charged AdS2 black holes were explored. Charged black holes with scalar fields challenge the no-hair theorem, affecting their entropy, temperature, and phase transitions, as discussed in Ref. \cite{Blagojevic:2023kqj}. Furthermore, the Iyer-Wald formalism has been used to derive thermodynamic laws for AdS-Kerr black holes, addressing controversies in the field, as studied in Ref. \cite{Xiao:2023lap}. The impact of topological properties on black hole thermodynamic stability has also been considered \cite{Wu:2022whe,Wei:2021vdx,Rizwan:2023ivp}. Critical points of charged Gauss-Bonnet black holes in AdS were classified in Ref. \cite{Yerra:2022alz}. Additional studies include the tunneling of charged bosons from noncommutative charged black holes, from which the rate and Hawking temperature were derived \cite{Ovgun:2015box}, the formulation of thermodynamics and computation of dual stress-energy tensors for accelerating black holes in AdS \cite{Anabalon}, and the generalization of non-relativistic black hole solutions to include electric charge, exploring their properties \cite{Jusufi}.

In summary, black hole thermodynamics not only enhances our understanding of the fundamental properties of black holes but also plays a central role in the quest for a theory of quantum gravity. Through the study of black hole entropy, temperature, and phase transitions, physicists are uncovering profound connections between gravity, quantum mechanics, and thermodynamics. As research continues to evolve, black hole thermodynamics remains a fertile ground for exploring some of the deepest questions in theoretical physics.}

In addition to the thermodynamic properties of black holes, their quasinormal modes (QNMs) and greybody factors are critical for understanding the dynamics and observational signatures of these enigmatic objects. Quasinormal modes represent the characteristic oscillations that black holes undergo when they are perturbed \cite{Konoplya:2003ii, Lambiase:2024lvo,Konoplya:2023ppx, Konoplya:2023aph, Konoplya:2022hll, Konoplya:2019hlu}. These oscillations are damped over time due to the emission of gravitational waves, and the frequencies and damping rates of the QNMs are determined by the black hole’s mass, charge, and angular momentum. The study of QNMs provides invaluable insights into the stability of black holes, as well as their interactions with external perturbations. Furthermore, QNMs are directly related to the ringdown phase of gravitational wave signals detected by observatories such as LIGO and Virgo. By analyzing the QNMs from observed gravitational waves, scientists can infer the parameters of the black hole, offering a unique observational tool for testing general relativity in the strong-field regime and exploring potential deviations from it in modified theories of gravity.

Closely related to QNMs are greybody factors, which describe the modification of Hawking radiation as it escapes from the black hole’s gravitational well. In the absence of any scattering, Hawking radiation would have a perfect blackbody spectrum, but greybody factors account for the fact that some of the radiation is scattered back into the black hole by the spacetime curvature. This scattering alters the spectrum of the emitted radiation and provides important corrections to the pure Hawking radiation spectrum. The study of greybody factors is crucial for understanding the energy emission rates of black holes and for exploring potential observational signatures of black holes in the universe. Together with QNMs, greybody factors enrich our understanding of black hole thermodynamics and play a key role in connecting theoretical predictions with astrophysical observations, particularly in the era of gravitational wave astronomy.

In this paper, we adopt the metric signature $(- , +, +, +)$ and use geometrical units where $c = G = \hbar = 1$. After this brief introduction, the paper is structured as follows: In Section \ref{Sec2}, we provide a concise overview of the deformed black hole solution. Section \ref{Sec3} examines the Hawking temperature of the black hole and explores the remnant radius and mass. In Section \ref{Sec4}, we analyze the photon sphere radius based on the motion of light and derive the black hole's shadow radius. Section \ref{Sec5} investigates the quasinormal modes (QNMs) of the black hole spacetime, while Section \ref{Sec6} focuses on the greybody factors. The emission rate of the black hole is discussed in Section \ref{Sec7}. { The topological characteristics of the black hole is studied in Section \ref{Sec9}.} Finally, in Section \ref{Sec8}, we conclude with a brief summary and discussion.

\section{Deformed Black hole solution}\label{Sec2}

In this section, we explore the construction of a deformed Anti-de Sitter (AdS) black hole solution within a modified gravity framework. The action governing the four-dimensional spacetime is given by \cite{Khosravipoor}

\begin{equation}
    A = \int d^4x \, \sqrt{-g} \left( \frac{R - 2\Lambda}{2\kappa} + \mathcal{L}_m + \mathcal{L}_{\rm X} \right),
\end{equation}
where $ R $ is the Ricci scalar, $ \Lambda $ is the cosmological constant, $ \kappa = 8\pi G/c^4 $, and $ \mathcal{L}_m $ and $ \mathcal{L}_{\rm X} $ represent the matter and beyond-General Relativity (GR) Lagrangians, respectively. The term $ \mathcal{L}_{\rm X} $ encompasses contributions from extra fields such as scalar, vector, or tensor fields, or extensions of gravity theories that modify the standard Einstein-Hilbert action. When $ \mathcal{L}_{\rm X} = 0 $, the resulting solution corresponds to the familiar AdS-Schwarzschild metric. However, the presence of $ \mathcal{L}_{\rm X} $ leads to a deformed black hole solution with interesting physical implications.

The Einstein field equations arising from this action can be written as

\begin{equation}
    G_{\mu \nu} + \Lambda g_{\mu \nu} = \kappa \, T^{(\rm tot)}_{\mu \nu},
\end{equation}
where $ G_{\mu \nu} $ is the Einstein tensor, and the total stress-energy tensor $ T^{(\rm tot)}_{\mu \nu} $ includes contributions from both the matter content and the extra fields described by $ \mathcal{L}_{\rm X} $, i.e.,

\begin{equation}
    T^{(\rm tot)}_{\mu \nu} = T^{(\rm m)}_{\mu \nu} + T^{(\rm X)}_{\mu \nu}.
\end{equation}

To construct the deformed black hole solution, we consider the absence of matter field and a static, spherically symmetric spacetime with the line element

\begin{equation} \label{final_line_element}
    ds^2 = -\mathcal{F}(r) dt^2 + \dfrac{1}{\mathcal{F}(r)} dr^2 + r^2 (d\theta^2 + \sin^2{\theta} \, d\phi^2).
\end{equation}
 The deformation in the black hole geometry is introduced via an energy density function $ \mathcal{E}(r) $, which is chosen to be proportional to $ 1/r^4 $, as follows \cite{Khosravipoor}:

\begin{equation} \label{eq5}
     \mathcal{E}(r) = \frac{\alpha}{\kappa (\beta + r)^4},
\end{equation}
where $ \alpha $ and $ \beta $ are deformation parameters. The parameter $ \alpha $ controls the strength of the deformation, while $ \beta $ ensures the avoidance of central singularities at $ r = 0 $. This choice of energy density function satisfies the necessary energy conditions and asymptotically decays as $ r \to \infty $. { The specific form of $\mathcal{E}(r)$ in Eq. \eqref{eq5} is motivated by the requirement to construct a regular black hole geometry that avoids curvature singularities at the centre. The $(\beta + r)^4$ dependence ensures that the energy density remains finite as $r \to 0$ and falls off rapidly as $r \to \infty$, thereby satisfying the necessary energy conditions and preserving the asymptotic (A)dS structure. }

Substituting $ \mathcal{E}(r) $ into the field equations yields the modified metric function for the deformed AdS black hole \cite{Khosravipoor}:

\begin{equation}\label{metric function}
 \mathcal{F}(r)=1-\frac{2M}{r}-\frac{\Lambda r^2}{3} +\alpha\,\frac{\beta^2+3r^2+3\beta r}{3r(\beta+r)^3}.
\end{equation}
where $ M $ is the mass of the black hole. The asymptotic form of the metric function remains consistent with the AdS-Schwarzschild solution, but the additional term proportional to $ \alpha $ introduces deviations that modify the black hole's properties.

The deformed AdS black hole solution offers a rich framework for exploring deviations from GR in strong gravitational regimes. The modifications introduced by $ \alpha $ and $ \beta $ affect key aspects of black hole physics, such as quasinormal modes (QNMs), greybody factors, and gravitational wave emission during black hole mergers. The variations in the black hole’s response to perturbations, particularly in the ringdown phase of mergers, provide potential observational signatures that could help distinguish deformed black holes from their standard counterparts.

Moreover, the inclusion of extra fields or modifications to the gravitational sector, as encapsulated by $ \mathcal{L}_{\rm X} $, aligns with ongoing efforts to reconcile GR with quantum gravity and other beyond-standard-model physics. By examining the behaviour of the deformed black hole in both classical and quantum contexts, we can gain deeper insights into the nature of gravity and the possible extensions of Einstein’s theory.

In the following sections, we will investigate the quasinormal modes and greybody factors associated with this deformed black hole solution and explore their observational consequences.

\section{Temperature}\label{Sec3}
\begin{figure}[htbp]
\centerline{
   \includegraphics[scale = 0.48]{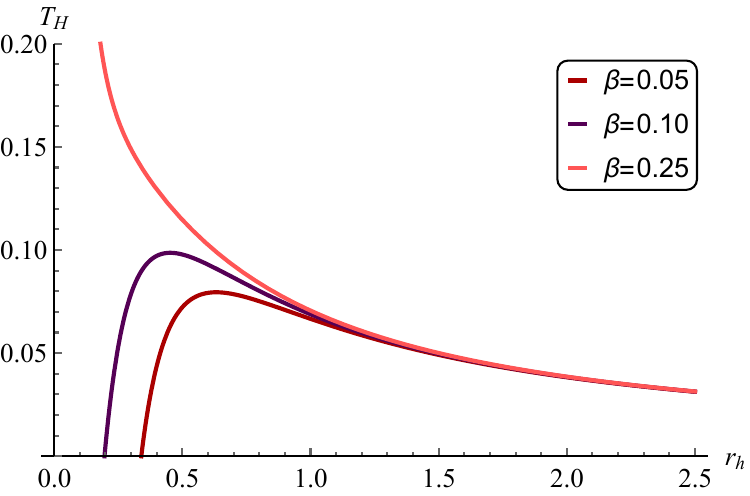}\hspace{0.5cm}
   \includegraphics[scale = 0.48]{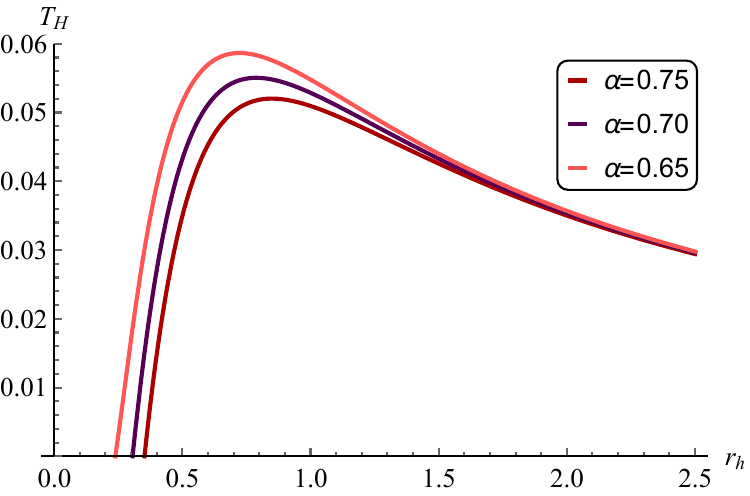}} \vspace{-0.2cm}
\caption{Variation of the Hawking temperature with radial distance for $\beta=0.2,\,\Lambda=-0.002$ (right panel) and $\alpha=0.2,\,\Lambda=-0.002$ (left panel) respectively.}
\label{fig:TH}
\end{figure}

\begin{figure}[htbp]
\centerline{
   \includegraphics[scale = 0.48]{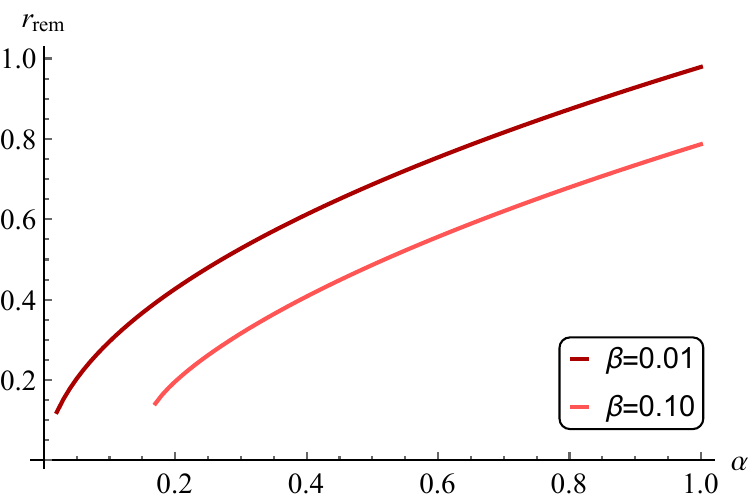}} \vspace{-0.2cm}
\caption{Behaviour of remnant radius as a function of parameter $\alpha$ with $\Lambda=-0.002$.}
\label{fig:rRem}
\end{figure}

\begin{figure}[ht!]
  \includegraphics[width=6.5cm]{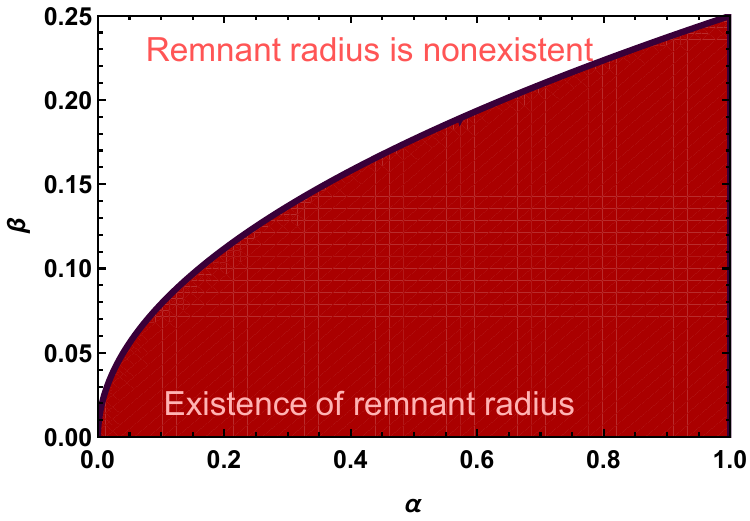} \hspace{-0.2cm}\\
    \caption{Region where remnant radius does exist, assuming $\Lambda=0.01$.}\label{fig:rem}
\end{figure}

At {the} horizon $\mathcal{F}(r)=0\big|_{r=r_h}$, thus one can find the mass of the black hole as a function of the horizon radius.
\begin{eqnarray}
M(r_h)=-\frac{1}{2} r_h \left(-\frac{\alpha  \left(\beta ^2+3 r_h^2+3 \beta  r_h\right)}{3 r_h (\beta +r_h)^3}+\frac{\Lambda  r_h^2}{3}-1\right).\label{Mrh}
\end{eqnarray}
Hawking temperature is obtained from $\frac{1}{4\pi}\partial \mathcal{F}(r)/\partial r|_{r=r_h}$. For the metric of the form \ref{metric function}, by substituting the mass from Eq. \eqref{Mrh}, the temperature would be found as
\begin{eqnarray}
T_H={\frac{\beta ^4+r_h^4+4 \beta  r_h^3-\alpha  r_h^2+6 \beta ^2 r_h^2-\Lambda  r_h^2 (\beta +r_h)^4+4 \beta ^3 r_h}{4 \pi  r_h (\beta +r_h)^4}}\label{TH}
\end{eqnarray}
The temperature curve is illustrated in Fig. \ref{fig:TH}.
\\We are interested in cases that temperature curve crosses $r_h$ axis in a positive value of $r_h$, this point is called remnant radius and is calculated as
\begin{eqnarray}
T_H\Big|_{r=r_{rem}}=\frac{\beta ^4+r_{rem}^4+4 \beta  r_{rem}^3-\alpha  r_{rem}^2+6 \beta ^2 r_{rem}^2-\Lambda  r_{rem}^2 (\beta +r_{rem})^4+4 \beta ^3 r_{rem}}{4 \pi  r_{rem} (\beta +r_{rem})^4}=0\label{rRem}
\end{eqnarray}
Remnant radius as a function of parameter $\alpha$ is demonstrated in Fig. \ref{fig:rRem}
As is apparent, while with $\beta$, and $\Lambda$ held constant, an increase in the parameter $\alpha$ results in an enlargement of the remnant radius, with $\alpha$, and $\Lambda$ held constant, an increase in the parameter $\beta$ leads to a decrease in the value of the remnant radius.
\\{Fig. \ref{fig:rem} illustrates the allowable range for the parameters $\alpha$ and $\beta$, for which there is a remnant radius for a specific value of parameter $\Lambda$.
}
\\{By substituting the remnant radius obtained from Eq. \eqref{rRem}, by $r_h$ in Eq. \eqref{Mrh}, the remnant mass of the black hole can be expressed as follows}
\begin{eqnarray}
M(r_{rem})=-\frac{1}{2} r_{rem} \left(-\frac{\alpha  \left(\beta ^2+3 r_{rem}^2+3 \beta  r_{rem}\right)}{3 r_{rem} (\beta +r_{rem})^3}+\frac{\Lambda  r_{rem}^2}{3}-1\right).\label{Mrem}
\end{eqnarray}
\section{Photon and Shadow Radii}\label{Sec4}
\begin{figure}[htbp]
\centerline{
   \includegraphics[scale = 0.48]{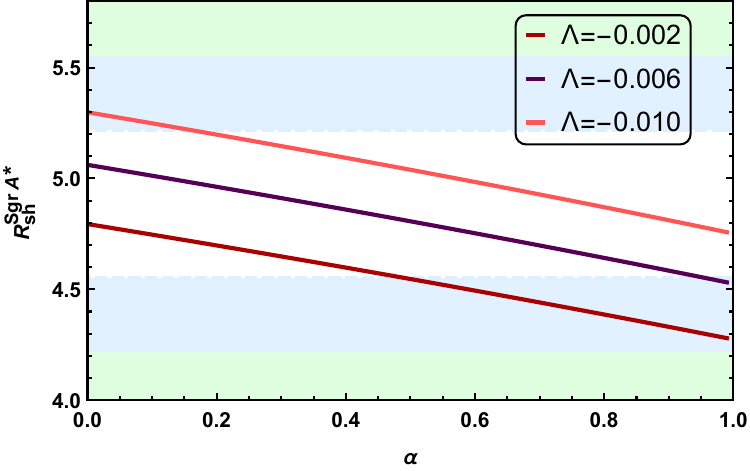}\hspace{0.5cm}
   \includegraphics[scale = 0.48]{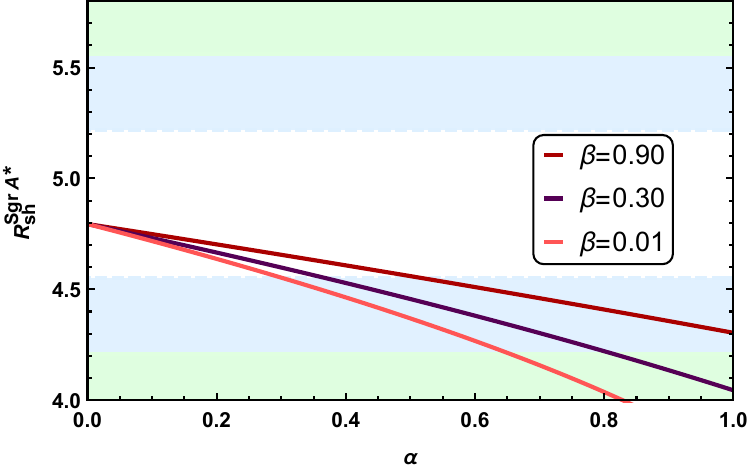}} \vspace{-0.2cm}
\caption{Variation of shadow radius with parameter $\alpha$ for $r_o = 10M$. In the left panel, $\ beta=0.8$ and in the right panel $\Lambda=-0.002$ are considered.}
\label{fig:RshAlpha}
\end{figure}

\begin{figure}[htbp]
\centerline{
   \includegraphics[scale = 0.48]{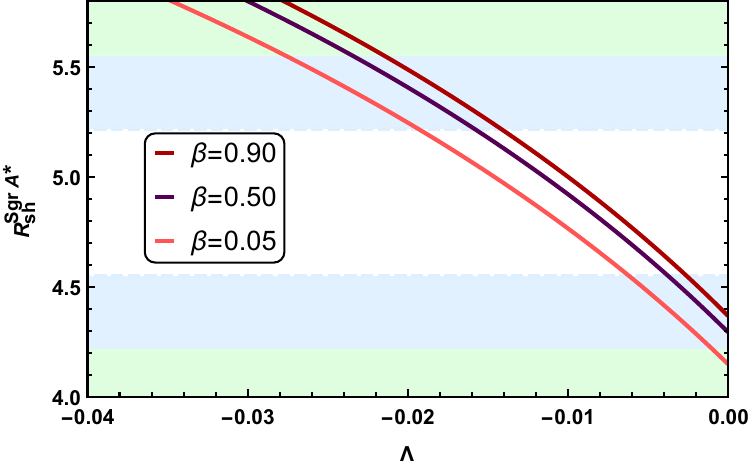}\hspace{0.5cm}
   \includegraphics[scale = 0.48]{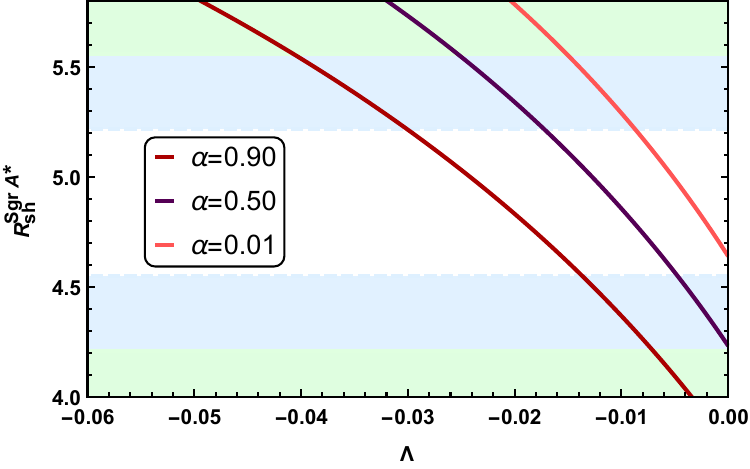}} \vspace{-0.2cm}
\caption{Variation of shadow radius with parameter $\Lambda$ for $r_o = 10M$. In the left panel $\alpha =0.6$ and in the right panel $\beta=0.02$ are considered.}
\label{fig:RshLambda}
\end{figure}

Assuming a massless case, the Euler-Lagrange equation is written as
\begin{eqnarray}
\frac{d}{d\tau}(\frac{\partial \mathcal{L}}{\partial \dot{x}^{\mu}})-\frac{\partial \mathcal{L}}{\partial x^{\mu}}=0\label{ELeq},
\end{eqnarray}
where the Lagrangian is obtained as \cite{Ahmed}
\begin{eqnarray}
\mathcal{L}=\frac{1}{2}g_{\mu\nu}\dot{x}^{\mu}\dot{x}^{\nu}.
\end{eqnarray}
metric tensor $g_{\mu\nu}$ can be calculated from Eq. \eqref{final_line_element}. In the equatorial plane where $\theta=\pi/2$, the Lagrangian yields
\begin{eqnarray}
2\mathcal{L}=-\mathcal{F}(r)\dot{t}^2+\frac{\dot{r}}{\mathcal{F}(r)}+r^2\dot{\phi}^2
\end{eqnarray}
Assuming two motion constants as $E=\partial\mathcal{L}/\partial\dot{t} =\mathcal{F}(r) \dot{t}$ and $L=\partial\mathcal{L}/\partial\dot{\phi} = r^2 \dot{\phi}$, where $E$ referred to the energy and $L$ indicates the angular momentum, for null geodesic Eq. \eqref{ELeq} reduces to
\begin{eqnarray}
\left(\frac{d r}{d \tau}\right)^2+\frac{L^2}{r^2}\mathcal{F}(r)-E^2=0.
\end{eqnarray}
Thus, one can define an effective potential which is formulated as
\begin{eqnarray}
V_{eff}(r)=\mathcal{F}(r)\frac{L^2}{r^2}-E^2
\end{eqnarray}
To ensure a circular orbit, two constraints are applied, which are as {follows}
\begin{eqnarray}
V_{Eff}(r)=0,\;\; \partial_r V_{Eff}(r)=0\label{CircularOrbit},
\end{eqnarray}
Thus, one can find that the following constraints should be satisfied
\begin{align}
E/L&=\pm\sqrt{\mathcal{F}(r)/r^2}\label{b},\\
r^2 \partial_r& \mathcal{F}(r)-2r \mathcal{F}(r)=0\label{rph}
\end{align}
By solving Eq. \eqref{rph} at $r=r_{ph}$, one can find the photon radius of the black hole.
\\Shadow radius is given by
\begin{eqnarray}
r_{sh}=r_{ph}\frac{\sqrt{f(r_o)}}{\sqrt{f(r_{ph})}},\label{rsh}
\end{eqnarray}
that $r_o$ indicates the observer radius.
Shadow radius is shown in Figs. \ref{fig:RshAlpha} and \ref{fig:RshLambda}. Table \ref{Table:rsh}, represents the permissible range of parameters based on the observation of Sgr A$^{*}$ for $1\sigma$ and $2\sigma$ regions.

\begin{figure}[htbp!]
\centerline{
   \includegraphics[scale = 0.48]{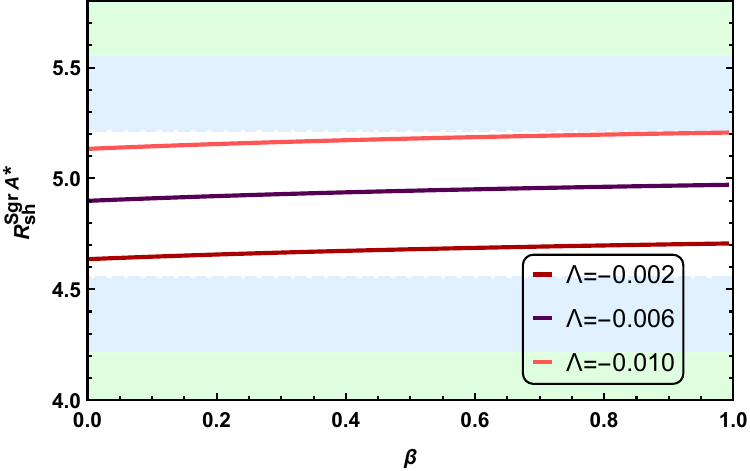}\hspace{0.5cm}
   \includegraphics[scale = 0.48]{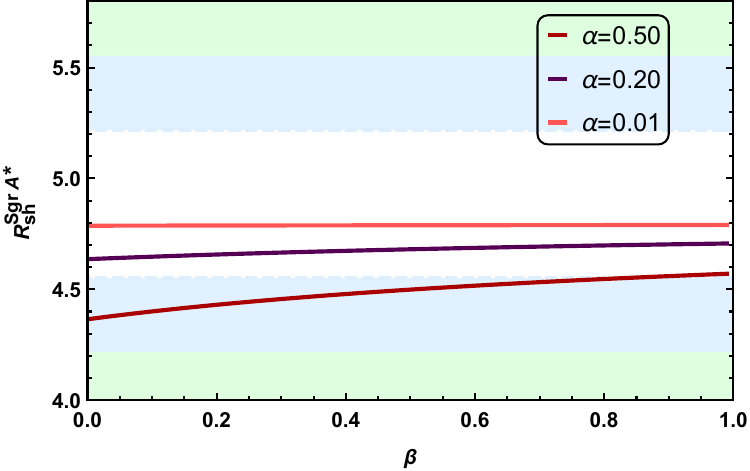}} \vspace{-0.2cm}
\caption{Variation of shadow radius with parameter $\beta$ for  $r_o = 10M$. In the left panel $\alpha =0.2$ and in the right panel $\Lambda=-0.002$ are considered.}
\label{fig:RshBeta}
\end{figure}
As evident in Figs. \ref{fig:RshAlpha} to \ref{fig:RshBeta}, a reduction in the parameters $\alpha$, and $\Lambda$, while the rest of the parameters remain constant, results in a decrease in the value of the shadow radius.

The allowed range of parameters $\alpha$ and $\Lambda$, where other parameters are constant, in which shadow radius lies in the $1\sigma$ and $2\sigma$ regions, due to the observations of Sgr A$^*$ is represented in Table \ref{Table:rsh}.
Also, as shown in Fig. \ref{fig:RshBeta}, by setting $\alpha=0.2$ and for the given $\Lambda$s, the value of shadow radius lies in the $1 \sigma$ region of Sgr A$^*$. By setting $\Lambda =-0.002$, for the case $\alpha =0.50$, for $\beta>0.83$, shadow radius is permissible in $1\sigma$ region due to the observations of Sgr A$^*$.
\begin{table}[htbp!]
		\caption{Upper bounds and lower bounds of the parameters based on the observations of Sgr A$^*$.}
		\centering
		\label{Table:rsh}
		\begin{tabular}{|ccccc|ccccc|}
		\hline
		\hline
		 & \multicolumn{2}{c}{$1\sigma$} & \multicolumn{2}{c|}{$2\sigma$} & & \multicolumn{2}{c}{$1\sigma$} & \multicolumn{2}{c|}{$2\sigma$}\\
		 \hline
		 $\alpha\,(\Lambda=-0.002)$ & Lower & Upper & Lower & Upper & $\alpha\,(\beta=0.800)$ & Lower & Upper & Lower & Upper\\
		  \hline
		 $\beta=0.900$ & $---$ & $---$ & $0.720$ & $---$ & $\Lambda=-0.002$ & $---$ & $---$ & $0.493$ & $---$\\
		 \hline
		 $\beta=0.030$ & $---$ & $---$ & $0.508$ & $0.813$ & $\Lambda=-0.006$ & $---$ & $---$ & $0.959$ & $---$\\
		 \hline
		 $\beta=0.010$ & $---$ & $---$ & $0.415$ & $0.653$ & $\Lambda=-0.010$ & $---$ & $0.156$ & $---$ & $---$\\
		 \hline
		 \hline
		 $\Lambda\,(\beta=0.020)$ & Lower& Upper& Lower & Upper & $\Lambda\,(\alpha=0.600)$ & Lower& Upper& Lower & Upper\\
		  \hline
		 $\alpha=0.900$ & $-0.041$ & $-0.030$ & $-0.014$ & $-0.007$ & $\beta=0.900$ & $-0.022$ & $-0.014$ & $-0.003$ & $---$\\
		 \hline
		 $\alpha=0.500$ & $-0.025$ & $-0.017$ & $-0.005$ & $---$ & $\beta=0.500$ & $-0.024$ & $-0.016$ & $-0.004$ & $---$\\
		 \hline
		 $\alpha=0.010$ & $-0.015$ & $-0.009$ & $---$ & $---$ & $\beta=0.050$ & $-0.028$ & $-0.019$ & $-0.006$ & $-0.001$\\
		 \hline
		 \hline
		\end{tabular}
	\end{table}

\section{Massless Scalar and Electromagnetic Perturbations in Black Hole Spacetimes}\label{Sec5}

In this section, we delve into the dynamics of massless scalar and electromagnetic perturbations within black hole spacetimes. Our analysis operates under the critical assumption that the perturbing fields, whether scalar or electromagnetic, exert a negligible influence on the black hole's background geometry. This assumption allows us to model the black hole spacetime as a fixed, unperturbed background, focusing instead on the behaviour of the perturbations themselves. This approach is instrumental in understanding the stability of black holes and the nature of the gravitational waves emitted when these astrophysical objects are disturbed.

To investigate these perturbations, we start by deriving Schr\"odinger-like wave equations of the Klein-Gordon type, which govern the behaviour of massless scalar fields in curved spacetime. These equations are derived from the conservation laws inherent in the black hole's spacetime, ensuring that the symmetries of the black hole's geometry are preserved. The resulting wave equations describe how the scalar field evolves as it interacts with the gravitational potential of the black hole, providing a foundational framework for calculating the quasinormal modes (QNMs) associated with these perturbations.

QNMs represent the characteristic oscillations that dominate the response of a black hole to external perturbations. These modes are defined by complex frequencies, which encapsulate both the oscillation frequency and the damping rate of the perturbations. To compute these QNMs, we utilize the higher-order WKB (Wentzel-Kramers-Brillouin) approximation method. The WKB method is a semi-analytical approach that is particularly well-suited for analyzing the complex potentials that arise in black hole spacetimes. It allows for the precise calculation of quasinormal frequencies, especially in the regime where the wavelength of the perturbation is comparable to the size of the black hole's event horizon.

Understanding the behaviour of QNMs in various black hole spacetimes, including those with modified gravity or additional fields, can enhance our understanding of black holes' fundamental properties and their role in gravitational wave astronomy. Here we specifically examine massless scalar and vector perturbations within the deformed black hole metric, as described by Eq.\eqref{final_line_element}. Under the assumption that the test fields (whether scalar or vector) have a negligible impact on the black hole's spacetime, we infer that any back-reaction on the metric is insignificant \cite{lopez2020, Chandrasekhar:1985kt}. To extract the QNMs, we derive Schrödinger-like wave equations for each type of perturbation by applying the relevant conservation laws associated with the spacetime under consideration. The resulting equations are of the Klein-Gordon type for scalar fields and the Maxwell type for electromagnetic fields.

To compute the QNMs, we employ the Pad\'e-averaged 6th-order WKB approximation method. This method provides a suitable framework, allowing for a more comprehensive analysis of the quasinormal spectra associated with different types of perturbations \cite{Gogoi:2023kjt, Gogoi:2024vcx, Gogoi:2023lvw}.

Focusing on axial perturbations, we express the perturbed metric in the form:
\begin{equation}  
\label{pert_metric}
ds^2 = -g_{tt} dt^2 + r^2 \sin^2\!\theta\, (d\phi - p_1(t,r,\theta)\, dt - p_2(t,r,\theta)\, dr - p_3(t,r,\theta)\, d\theta)^2 + g_{rr}\, dr^2 + r^2 d\theta^2,
\end{equation}
where the functions $p_1$, $p_2$, and $p_3$ represent the perturbations introduced into the black hole spacetime. The metric components $g_{tt} = \mathcal{F}(r)$ and $g_{rr} = 1/\mathcal{F}(r)$ correspond to the unperturbed, static, and spherically symmetric components in the perturbative expansion relative to the perturbations $p_i$.

By analysing these perturbations within the deformed black hole framework, we aim to further elucidate the stability and dynamical behaviour of black holes under scalar and electromagnetic disturbances. The insights gained from this analysis are expected to contribute to our broader understanding of black hole physics and the role of QNMs in gravitational wave astronomy.

We start by taking into account a massless scalar field in the vicinity of the black hole defined in \eqref{metric function}. Since the massless scalar $\Phi$ is assumed to have a negligible back reaction, the stationary Schr\"odinger equation
\begin{equation}  \label{radial_scalar}
\partial^2_{r_*} \psi(r_*)_{sl} + \omega^2 \psi(r_*)_{sl} = \mathcal{V}_s(r) \psi(r_*)_{sl},
\end{equation}
after defining the tortoise coordinate $r_*$ as 
\begin{equation}  \label{tortoise}
\dfrac{dr_*}{dr} = \sqrt{g_{rr} g_{tt}^{-1}} = \mathcal{F}(r)
\end{equation}
such that $\mathcal{V}_s(r)$ stands for the effective potential  \cite{lopez2020}
\begin{equation}  \label{Vs}
\mathcal{V}_s(r) = |g_{tt}| \left( \dfrac{l(l+1)}{r^2} +\dfrac{1}{r \sqrt{|g_{tt}|
g_{rr}}} \dfrac{d}{dr}\sqrt{|g_{tt}| g_{rr}^{-1}} \right) = \mathcal{F}(r) \left( \dfrac{l(l+1)}{r^2} +\dfrac{1}{r} \dfrac{d}{dr}\mathcal{F}(r) \right)
\end{equation}
so that, now, $l$ becomes the multipole moment of the black hole's
QNMs.
The frequency $\omega$ in the stationary Schr\"odinger equation \eqref{radial_scalar} is defined via $\partial_t^2 \psi_{sl} (t,r) =-\omega^2 \psi_{sl}(t,r)$ and this definition is made possible by the diagonal and static nature of the metric in \eqref{final_line_element}. The wave function $\psi(r_*)_s$ stands for the waves associated with the scalar QNMs \cite{Chandrasekhar:1985kt,lopez2020}. 
{The maximum of this potential is located at $\dfrac{\partial}{\partial r}\mathcal{V}_s(r)\bigg|_{r=r_{peak}}=0$.}

Having done with the scalar perturbations, now we move to the massless vector perturbations, which are nothing but electromagnetic perturbations. In analysing vector perturbations it proves useful to use the tetrad formalism  \cite{Chandrasekhar:1985kt,lopez2020} in which the curved metric $g_{\mu\nu}$ in \eqref{final_line_element} is projected on the flat metric $\eta_{\bar \mu \bar \nu}$ via the vierbein $e_\mu^{\bar \mu}$ in the form $g_{\mu\nu}= e_\mu^{\bar \mu}\eta_{\bar \mu \bar \nu}e^{\bar \nu}_\nu$. (Here, the notation is such that while $\mu, \nu, \dots$ refer to curved spacetime, $\bar \mu,  \bar \nu, \dots$ refer to flat spacetime.) Vector perturbations obey the stationary Schr\"odinger equation
\begin{equation}
\partial^2_{r_*} \psi_{el}(r_*) + \omega^2 \psi_{el}(r_*) = \mathcal{V}_{em}(r) \psi_{el}(r_*),
\end{equation}
with the effective potential 
\begin{equation}  \label{Ve}
\mathcal{V}_{em}(r) = g_{tt}\, \dfrac{l(l+1)}{r^2}=\mathcal{F}(r) \dfrac{l(l+1)}{r^2}\,.
\end{equation}
This effective potential $\mathcal{V}_{em}(r)$ of vector perturbations differs from the effective potential $\mathcal{V}_s(r)$ of scalar perturbations by the presence of the derivative of $g_{tt}$ in the latter{, and the peak of the potential $\mathcal{V}_{em}(r)$ is located at $\dfrac{\partial}{\partial r}\mathcal{V}_{em}(r)\bigg|_{r=r_{peak}}=0$}.

\begin{figure}[htbp]
\centerline{
   \includegraphics[scale = 0.8]{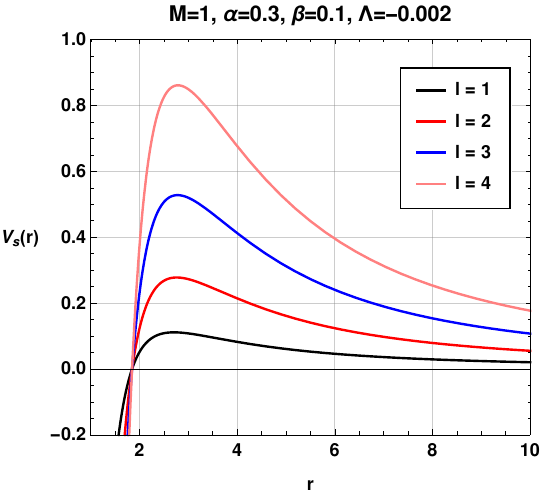}\hspace{0.5cm}
   \includegraphics[scale = 0.8]{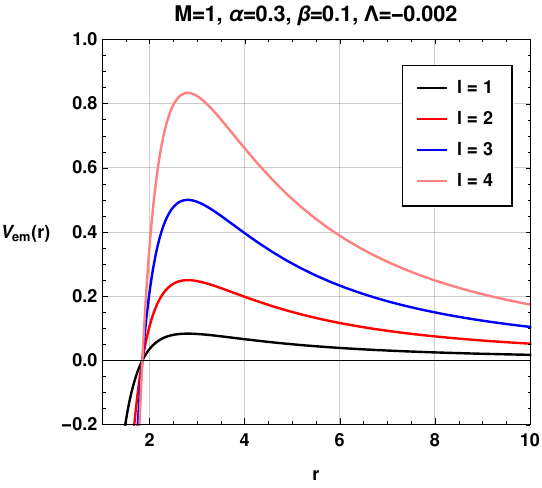}} \vspace{-0.2cm}
   
\caption{Behaviour of potentials for scalar and electromagnetic perturbations with $r$.}
\label{V01}
\end{figure}

\begin{figure}[htbp]
\centerline{
   \includegraphics[scale = 0.8]{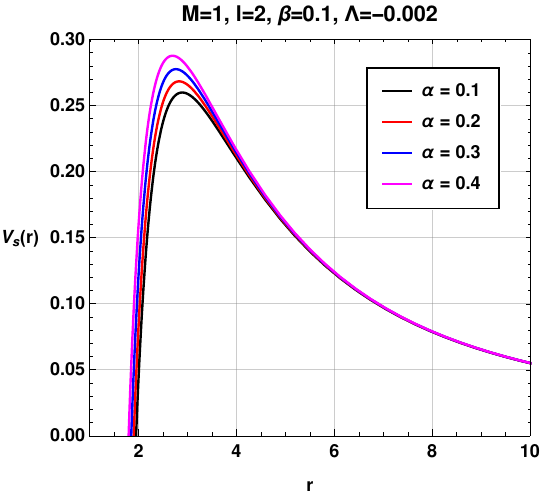}\hspace{0.5cm}
   \includegraphics[scale = 0.8]{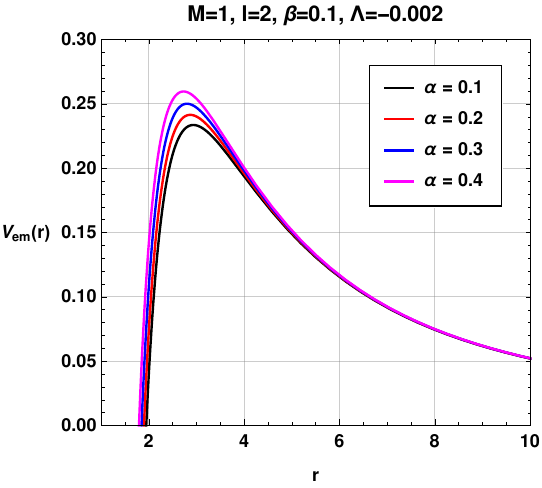}} \vspace{-0.2cm}
   
\caption{Behaviour of potentials for scalar and electromagnetic perturbations with $r$.}
\label{V02}
\end{figure}

\begin{figure}[htbp]
\centerline{
   \includegraphics[scale = 0.8]{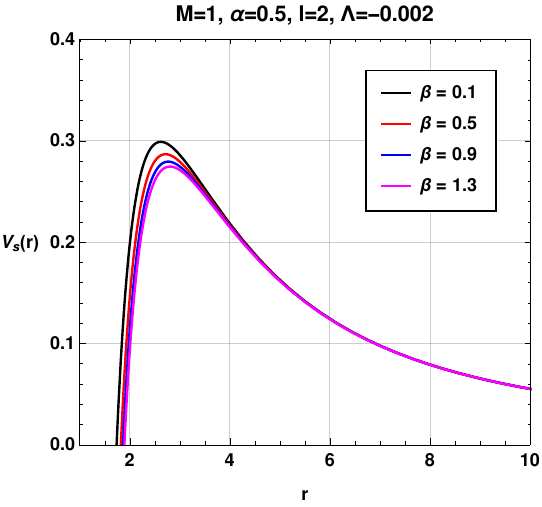}\hspace{0.5cm}
   \includegraphics[scale = 0.8]{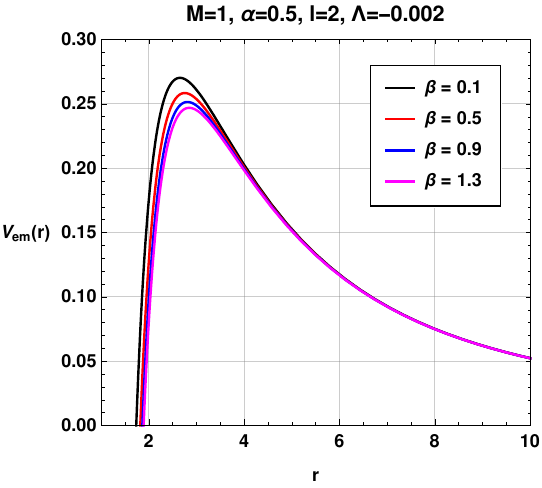}} \vspace{-0.2cm}
   
\caption{Behaviour of potentials for scalar and electromagnetic perturbations with $r$.}
\label{V03}
\end{figure}

In Fig.s \ref{V01}, \ref{V02} and \ref{V03}, we have depicted the behaviours of the scalar and electromagnetic potentials with respect to the model parameters $l$, $\alpha$ and $\beta$ respectively. From Fig. \ref{V01}, we can see that the potentials associated with both scalar and electromagnetic potentials behave similarly. However, for the scalar perturbation, the value of the peak of the potential is higher than that of the electromagnetic potential.

Fig. \ref{V02} depicts that with an increase in the value of model parameter $\alpha$, the peak value of the potential increases and the peak shifts towards the event horizon of the black hole for both types of perturbations. Finally, in Fig. \ref{V03}, we have seen that model parameter $\beta$ has an opposite impact on the peak of the potential in comparison to that of the model parameter $\alpha$. So, by investigating the behaviour of the scalar and electromagnetic potentials associated with the black hole spacetime, one can see that both the model parameters $\alpha$ and $\beta$ might have opposite impacts on the QNMs spectrum of the black hole. To understand the impacts of these model parameters on ring-down GWs, we now move to the investigation of QNMs in detail.

\begin{figure}[htbp]
\centerline{
   \includegraphics[scale = 0.6]{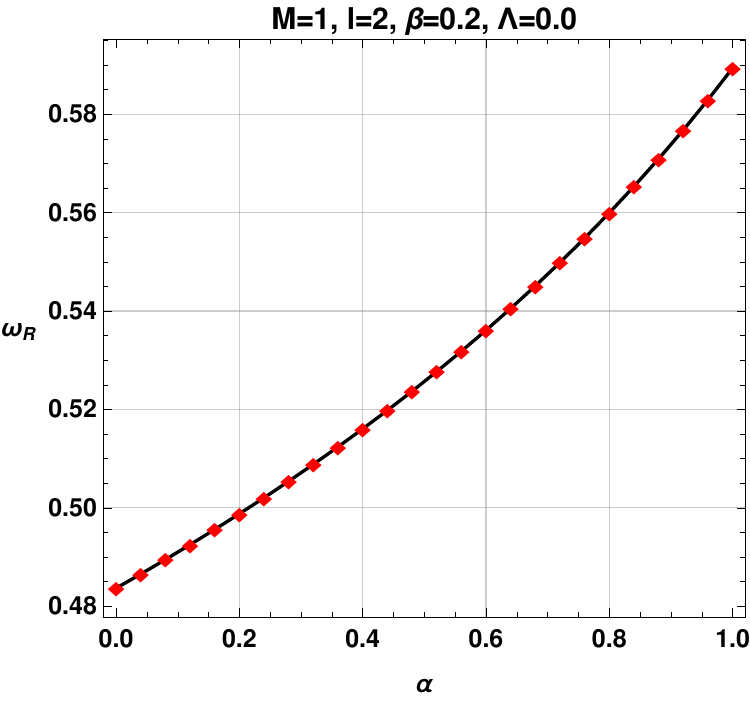}\hspace{0.5cm}
   \includegraphics[scale = 0.6]{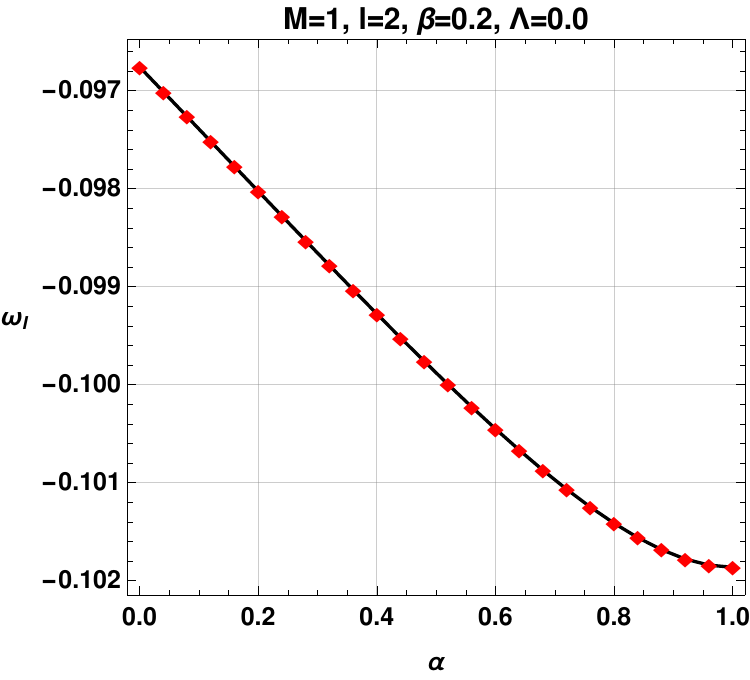}} \vspace{-0.2cm}
   
\caption{The variation of QNMs for scalar perturbation with $\alpha$.}
\label{QNM01}
\end{figure}

\begin{figure}[htbp]
\centerline{
   \includegraphics[scale = 0.6]{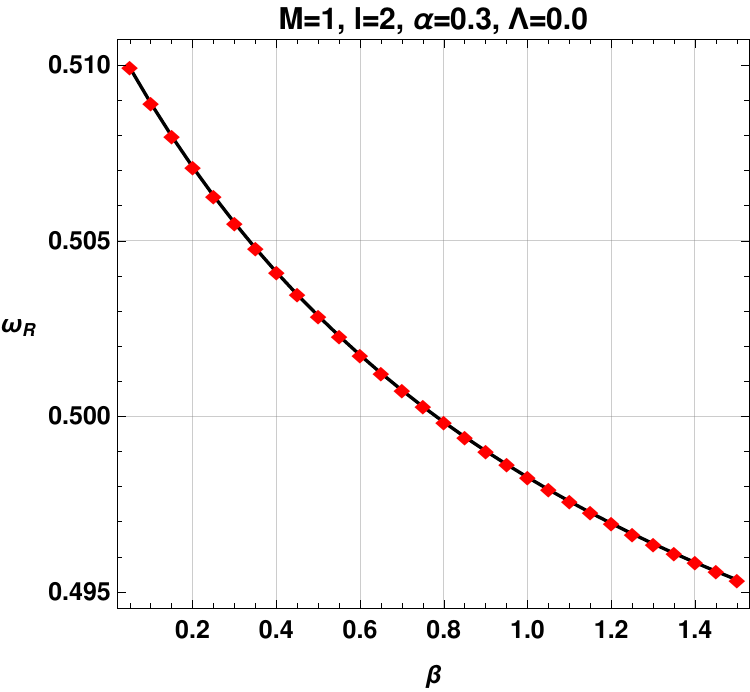}\hspace{0.5cm}
   \includegraphics[scale = 0.6]{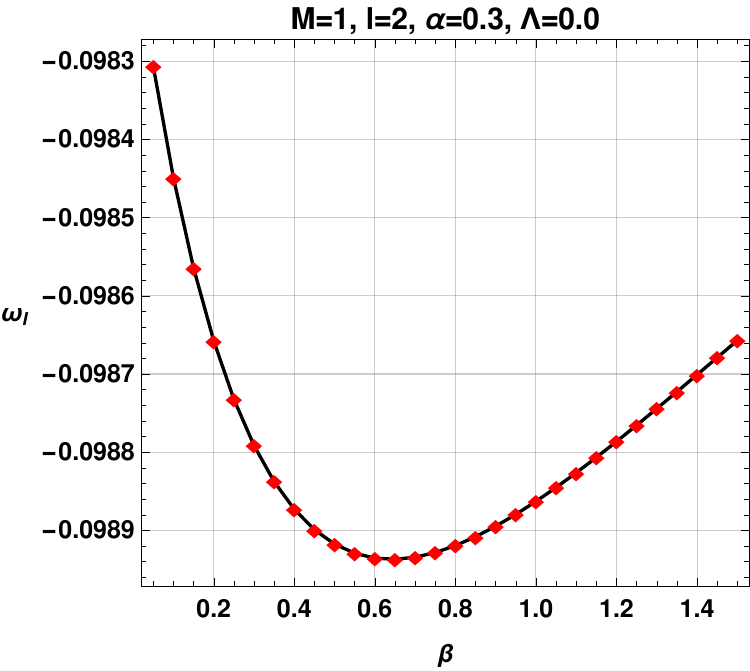}} \vspace{-0.2cm}
   
\caption{The variation of QNMs for scalar perturbation with $\beta$.}
\label{QNM02}
\end{figure}

\begin{figure}[htbp]
\centerline{
   \includegraphics[scale = 0.6]{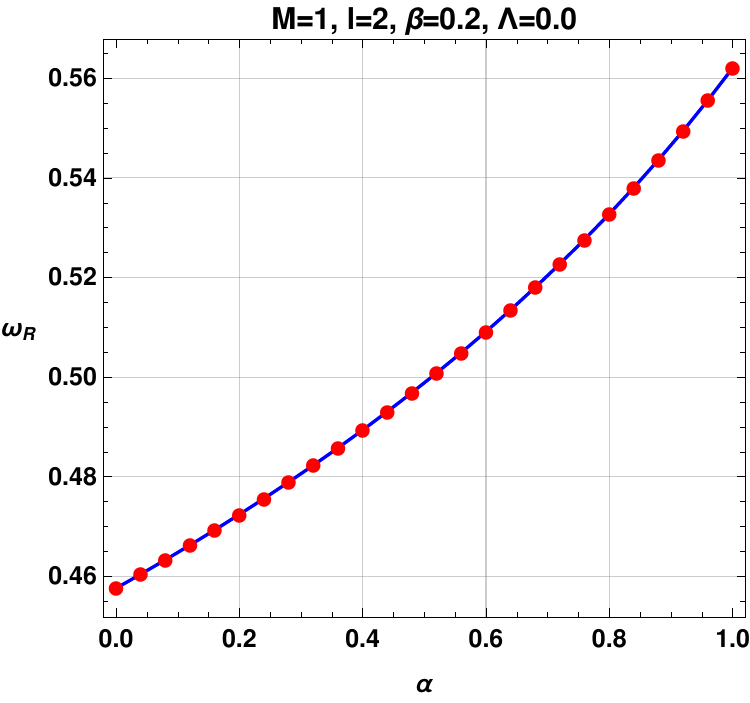}\hspace{0.5cm}
   \includegraphics[scale = 0.6]{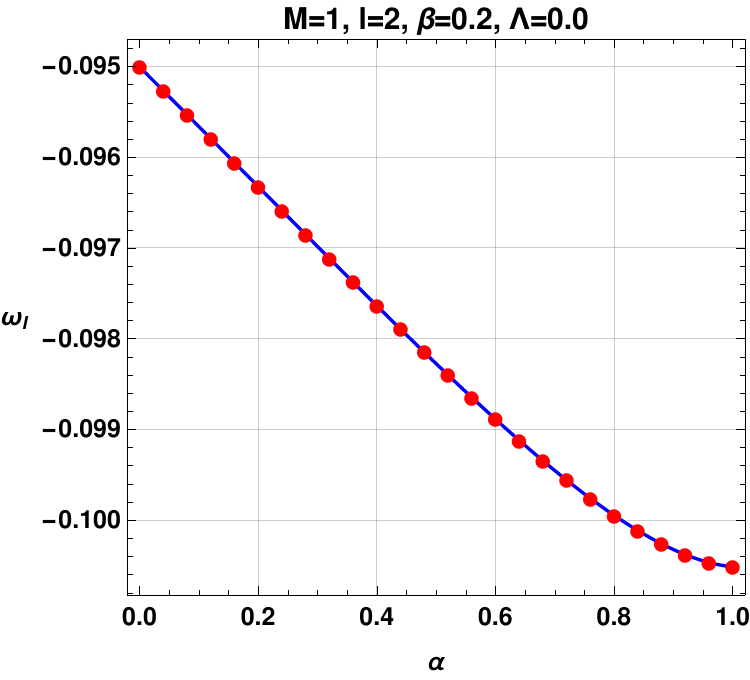}} \vspace{-0.2cm}
   
\caption{The variation of QNMs for electromagnetic perturbation with $\alpha$.}
\label{QNM03}
\end{figure}

\begin{figure}[htbp]
\centerline{
   \includegraphics[scale = 0.6]{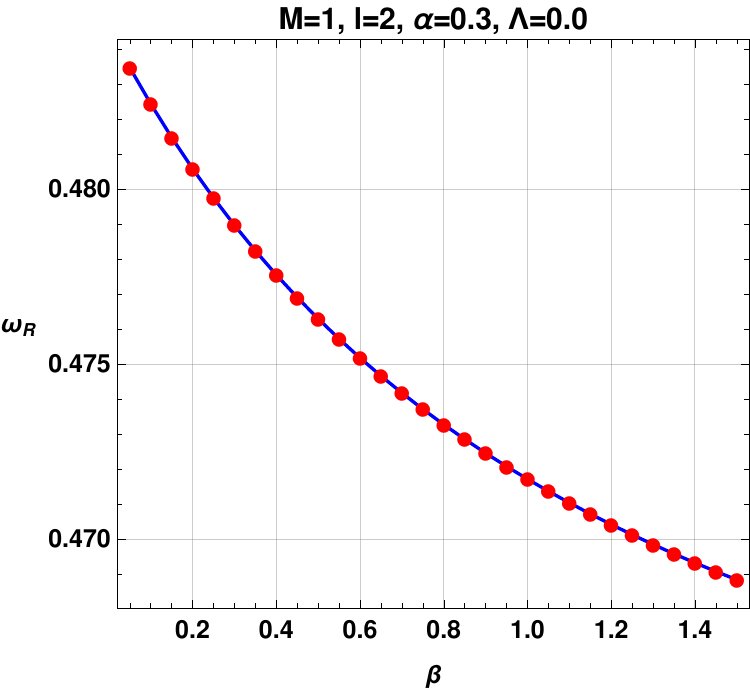}\hspace{0.5cm}
   \includegraphics[scale = 0.6]{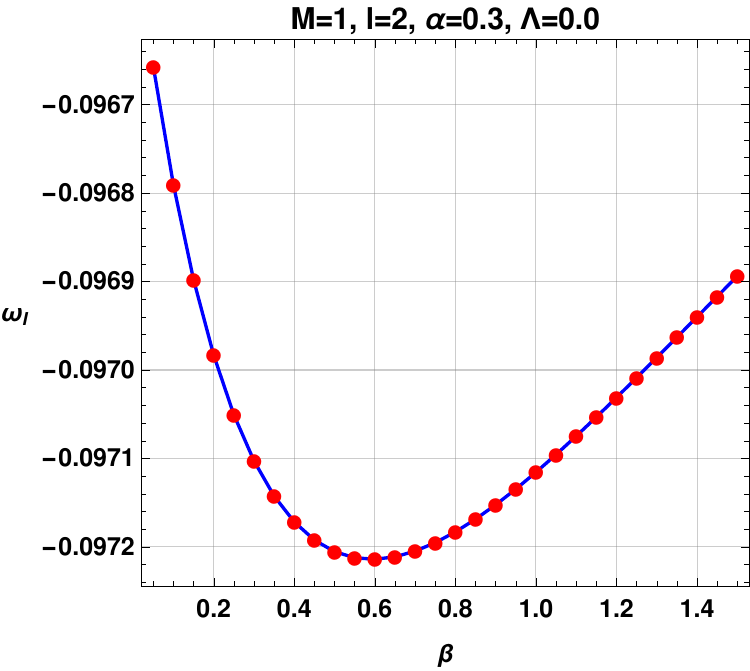}} \vspace{-0.2cm}
   
\caption{The variation of QNM for electromagnetic perturbation with $\beta$.}
\label{QNM04}
\end{figure}

This investigation employed the 6th-order WKB approximation technique. Within this method, the oscillation frequency $\omega$ of GWs can be determined using the following expression:
\begin{equation}
\omega = \sqrt{-\, i \left[ (n + 1/2) + \sum_{k=2}^6 \bar{\Lambda}_k \right] \sqrt{-2 \mathcal{V}_0''} + \mathcal{V}_0},
\end{equation}
where $n = 0, 1, 2\hdots$, stands for overtone number, $\mathcal{V}_0 = \mathcal{V}|_{r\, =\, r_{max}}$ and 
$\mathcal{V}_0'' = \dfrac{d^2 \mathcal{V}}{dr^2}|_{r\, =\, r_{max}}$. The position $r_{max}$ corresponds to the location where the potential function $\mathcal{V}(r)$ attains its highest value. The correction terms $\bar{\Lambda}_k$ play a role in refining the accuracy of the calculations. The specific expressions for these correction terms, along with the Pad\'e averaging method, can be found in the
Ref.s \cite{Schutz,Iyer:1986np,Konoplya:2003ii,Matyjasek:2019eeu,Konoplya:2019hlu,Lambiase:2024lvo, Konoplya:2023ppx,Konoplya:2023aph,Konoplya:2022hll}.
These references provide detailed information regarding the mathematical forms of the correction terms and the procedure for Pad\'e averaging.

{ \begin{table}[ht]
\caption{The QNMs for the massless scalar and electromagnetic perturbation with $n= 0$, $M=1$, $\alpha = 0.3$, $\beta=0.2$ and $\Lambda = 0.0$ using the Pad\'e averaged 6th order WKB approximation method.}
\label{tab02}
\begin{center}
{\small 
\begin{tabular}{|c|c|c|c|c|c|c|}
\hline
$l$ & Scalar QNMs & EM QNMs & $\vartriangle_{rms}$(scalar) & $\vartriangle_{rms}$(em) & $\Delta_6$(scalar) & $\Delta_6$(em) \\ \hline
1 & $0.307201 - 0.0995284 i$ & $0.261742 - 0.0946083 i$ & $1.13985\times10^{-6}$ & $7.76923\times10^{-6}$ & $3.51816\times10^{-6}$ & $6.88485\times10^{-6}$ \\
2 & $0.507104 - 0.0986576 i$ & $0.480589 - 0.0969834 i$ & $3.78281\times10^{-7}$ & $5.01194\times10^{-7}$ & $2.01920\times10^{-6}$ & $5.55769\times10^{-7}$ \\
3 & $0.708088 - 0.0984074 i$ & $0.689286 - 0.0975634 i$ & $8.38391\times10^{-8}$ & $8.84296\times10^{-8}$ & $3.96363\times10^{-7}$ & $1.05988\times10^{-7}$ \\
4 & $0.909420 - 0.0983032 i$ & $0.894839 - 0.0977950 i$ & $2.42110\times10^{-8}$ & $2.46254\times10^{-8}$ & $1.13709\times10^{-7}$ & $3.05768\times10^{-8}$ \\
5 & $1.110910 - 0.0982502 i$ & $1.099000 - 0.0979107 i$ & $8.87193\times10^{-9}$ & $8.92710\times10^{-9}$ & $4.17001\times10^{-8}$ & $1.12966\times10^{-8}$ \\
6 & $1.312480 - 0.0982197 i$ & $1.302410 - 0.0979769 i$ & $3.83873\times10^{-9}$ & $3.83890\times10^{-9}$ & $4.90337\times10^{-9}$ & $4.92176\times10^{-9}$ \\
7 & $1.514110 - 0.0982005 i$ & $1.505380 - 0.0980182 i$ & $3.12285\times10^{-9}$ & $3.12212\times10^{-9}$ & $2.40862\times10^{-9}$ & $2.41337\times10^{-9}$ \\
8 & $1.715760 - 0.0981876 i$ & $1.708070 - 0.0980458 i$ & $1.47073\times10^{-9}$ & $1.47008\times10^{-9}$ & $1.29188\times10^{-9}$ & $1.29362\times10^{-9}$ \\
\hline
\end{tabular}
}
\end{center}
\end{table}
}

We have listed QNMs for $n=0$ and for different $l$ values in Table \ref{tab02} for both scalar and electromagnetic perturbations. These results were obtained for deformation parameters $ \alpha = 0.3 $ and $ \beta = 0.2 $, which influence the spacetime geometry and black hole properties. The table displays QNM frequencies, with the real part representing the oscillation frequency and the imaginary part corresponding to the damping rate of the modes. These modes were calculated for different multipole numbers $ l $, ranging from $ l = 1 $ to $ l = 8 $. { To ensure the reliability of our analysis, we have carefully chosen parameter values that maintain a well-defined, single-peak potential structure, which is a prerequisite for the WKB method to yield meaningful results. }

As expected, the real part of the QNMs increases with $ l $, indicating higher oscillation frequencies for modes with larger angular momentum. The imaginary part, which controls the damping or decay rate, shows slight variation across the different values of $ l $. Interestingly, scalar perturbations consistently exhibit slightly higher oscillation frequencies and more pronounced damping compared to electromagnetic perturbations. This is reflected in the larger imaginary components for scalar modes compared to their electromagnetic counterparts, highlighting the different behaviours of these two types of perturbations in deformed black hole spacetimes.

The table also presents the root mean square (rms) errors, $ \Delta_{rms} $, for both scalar and electromagnetic perturbations. These errors quantify the precision of the WKB method in estimating the QNMs. It is observed that the RMS errors generally decrease with increasing $ l $, indicating more accurate QNM calculations for higher multipole numbers. The WKB errors, $ \Delta_6 $, which {are} basically an error term associated with half of the absolute difference between QNMs calculated using 5th and 7th order WKB methods, similarly decrease with increasing $ l $. This suggests that the WKB method becomes more reliable for larger values of $ l $, where the approximation yields more precise QNM frequencies. Nevertheless, the electromagnetic perturbations exhibit slightly larger errors compared to scalar perturbations, although both remain within acceptable limits, particularly for $ l \geq 2 $.

In Fig. \ref{QNM01}, we have shown the variation of QNMs with respect to the model parameter $\alpha$ for scalar perturbations. On the first panel, we have shown the variations of the real part of QNMs i.e. the oscillation frequencies of ring-down GWs with respect to the model parameter $\alpha$. With an increase in the value of $\alpha$, oscillation frequency increases significantly. In the case of imaginary QNMs, we can see that an increase in the value of $\alpha$ increases the damping rate of ring-down GWs. In Fig. \ref{QNM02}, we have depicted the variations of QNMs with respect to the model parameter $\beta$. On the first panel of Fig. \ref{QNM02}, one can see that with an increase in the value of $\beta$, the real part of QNMs, i.e. oscillation frequency of GWs decreases non-linearly. In the case of damping rate, we observe an interesting behaviour. Initially, the damping rate of the decay rate of ring-down GWs increases with an increase in the value of $\beta$ up to around $\beta = 0.65.$ Beyond this value, the decay rate decreases with an increase in the model parameter $\beta$. In the cases of electromagnetic perturbations, we observe a similar scenario with slightly lower values of oscillation frequencies as shown in Fig.s \ref{QNM03} and \ref{QNM04}. { This feature arises due to the dual role of $\beta$, which not only ensures regularity at the core by avoiding the central singularity but also modifies the effective potential governing scalar perturbations. Initially, increasing $\beta$ enhances the damping (i.e., $\omega_I$ becomes more negative) due to a steeper potential barrier; however, beyond a critical value, the core regularization becomes dominant, leading to a broadening or softening of the potential that supports longer-lived modes, hence a reversal in the damping trend. This behaviour may signal a transition in the internal structure of the spacetime and could carry potential astrophysical implications if such geometries model exotic compact objects—offering observable signatures in gravitational wave ringdowns that deviate from classical black hole predictions.}

The results offer important insights into the physical behaviour of deformed AdS black holes, especially in relation to the parameters $ \alpha $ and $ \beta $, which modify the spacetime structure and consequently influence the QNM frequencies and damping rates. The distinct responses of the black hole to scalar and electromagnetic perturbations underscore the complexity of black hole interactions with different types of fields. These variations are critical for astrophysical phenomena such as gravitational wave emissions during black hole mergers, where the nature of the perturbing field can impact the ringdown phase. The reliability of the 6th-order WKB method is evident from its accuracy, particularly for higher multipole numbers, affirming its utility in these calculations. The intricate behaviour of the damping rates and oscillation frequencies, particularly the non-linear effects observed with the control parameter $ \beta $, suggests that further exploration of these deformation parameters within modified gravity frameworks could deepen our understanding of black hole stability, energy dissipation, and their broader astrophysical and theoretical significance.

{ Now, we would like to investigate the time domain profiles of the perturbations. To do so, we shall use the scheme introduced in Ref. \cite{gundlach}.
This methodology allows us to investigate the temporal evolution and decay characteristics of perturbations, particularly at late times, thereby furnishing an independent check on the stability and consistency of the system under consideration.

To implement this technique, we discretize the wave equation on a uniform grid by defining the scalar field as $\psi(r_*, t) = \psi(i \Delta r_*, j \Delta t) = \psi_{i,j}$, where $i$ and $j$ index the spatial and temporal grid points respectively. The effective potential is similarly discretized as $V(r(r_*)) = V(r_*, t) = V_{i,j}$, which in the time-independent case simplifies to $V_{i,j} = V_i$.

Using these discretized variables, the scalar wave equation takes the following finite difference form:
\begin{equation}
\frac{\psi_{i+1,j} - 2\psi_{i,j} + \psi_{i-1,j}}{\Delta r_*^2} - \frac{\psi_{i,j+1} - 2\psi_{i,j} + \psi_{i,j-1}}{\Delta t^2} - V_i \psi_{i,j} = 0.
\end{equation}
This equation can be explicitly solved for $\psi_{i,j+1}$, yielding the following iterative scheme for the time evolution of the perturbation:
\begin{equation}
\psi_{i,j+1} = -\psi_{i,j-1} + \left( \frac{\Delta t}{\Delta r_*} \right)^2 (\psi_{i+1,j} + \psi_{i-1,j}) + \left[ 2 - 2\left( \frac{\Delta t}{\Delta r_*} \right)^2 - V_i \Delta t^2 \right] \psi_{i,j}.
\end{equation}

The initial profile for the scalar perturbation is modelled as a Gaussian wave packet:
\begin{equation}
\psi(r_*, t=0) = \exp\left[-\frac{(r_* - k_1)^2}{2\sigma^2} \right],
\end{equation}
with the condition $\psi(r_*, t<0) = 0$, where $k_1$ and $\sigma$ denote the central location and width of the initial packet, respectively.

To ensure numerical stability of the iterative process, the ratio $\Delta t / \Delta r_*$ must satisfy the Von Neumann stability criterion, i.e.,
\begin{equation}
\frac{\Delta t}{\Delta r_*} < 1.
\end{equation}
This constraint ensures convergence of the solution and prevents the amplification of numerical errors during evolution.

By employing this finite-difference scheme, we numerically evolve the perturbation profile over time, thereby obtaining the time-domain signal. This evolution not only facilitates the identification of quasi-normal ringing but also serves as a crucial cross-validation for the quasi-normal mode spectrum derived in the frequency domain. Consequently, the combined analysis offers a more robust and comprehensive understanding of the perturbative dynamics in the considered gravitational background.

\begin{figure}[htbp]
\centerline{
   \includegraphics[scale = 0.68]{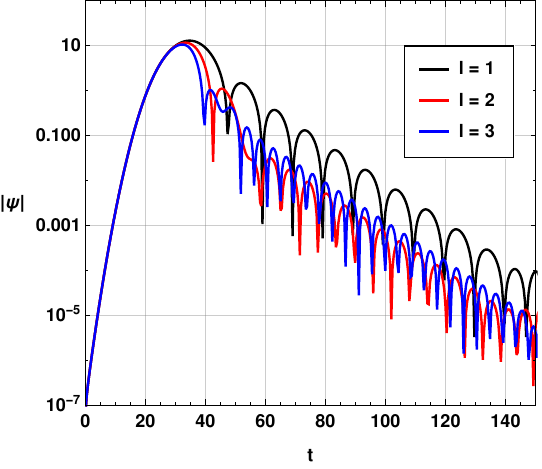}\hspace{0.5cm}
   \includegraphics[scale = 0.68]{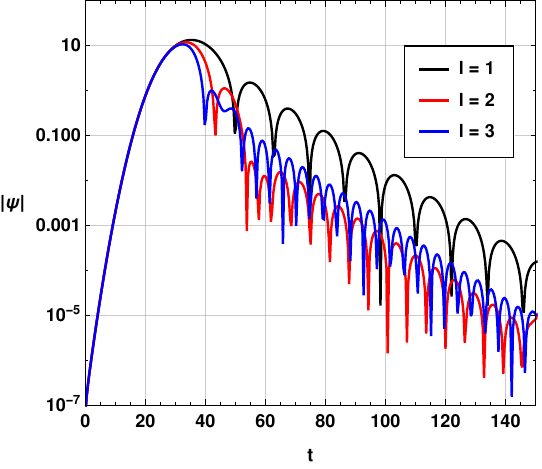}} 
   \vspace{-0.2cm}
\caption{The time-domain profiles of the massless scalar
perturbations (first panel) and electromagnetic perturbations (second panel) for different multipole moments $l$ with the parameter values $M=1$, $G = 1$, $n= 0$, $\alpha =0.3$, $\beta = 0.1$ and $\Lambda = -0.002$. }
\label{time01}
\end{figure}

\begin{figure}[htbp]
\centerline{
   \includegraphics[scale = 0.68]{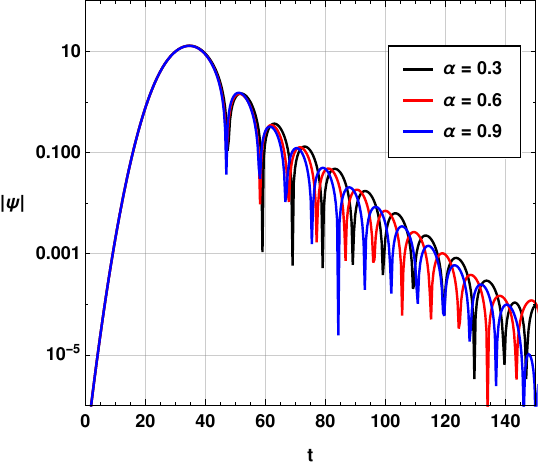}\hspace{0.5cm}
   \includegraphics[scale = 0.68]{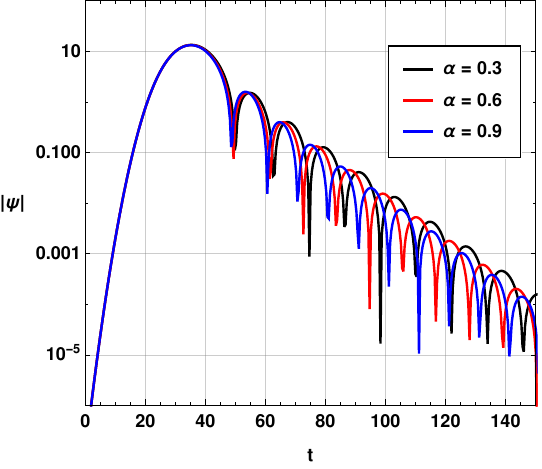}} 
   \vspace{-0.2cm}
\caption{The time-domain profiles of the massless scalar
perturbations (first panel) and electromagnetic perturbations (second panel) for different values of $\alpha$ with the parameter values $M=1$, $G = 1$, $n= 0$, $l =1$, $\beta = 0.1$ and $\Lambda = -0.002$. }
\label{time02}
\end{figure}

\begin{figure}[htbp]
\centerline{
   \includegraphics[scale = 0.68]{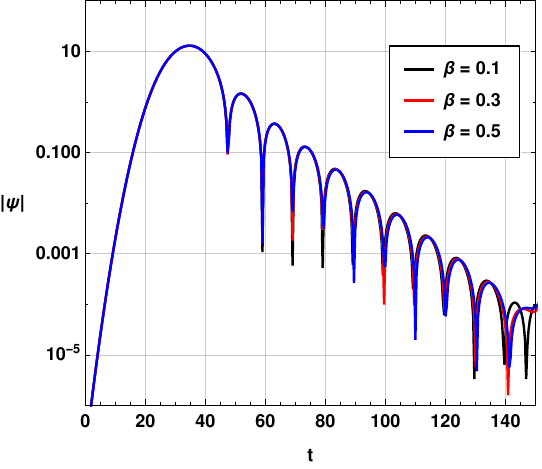}\hspace{0.5cm}
   \includegraphics[scale = 0.68]{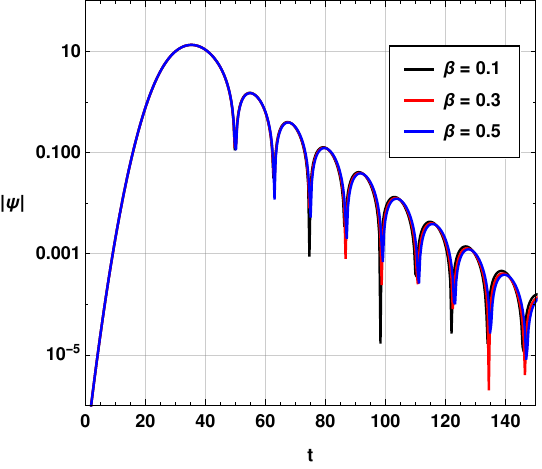}} 
   \vspace{-0.2cm}
\caption{The time-domain profiles of the massless scalar
perturbations (first panel) and electromagnetic perturbations (second panel) for different values of $\beta$ with the parameter values $M=1$, $G = 1$, $n= 0$, $\alpha =0.3$, $l = 1$ and $\Lambda = -0.002$. }
\label{time03}
\end{figure}

\begin{figure}[htbp]
\centerline{
   \includegraphics[scale = 0.68]{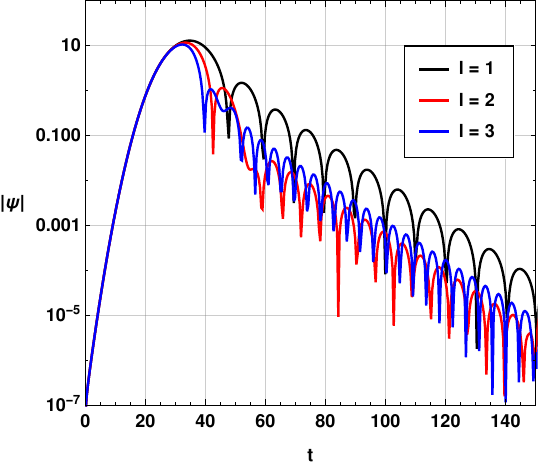}\hspace{0.5cm}
   \includegraphics[scale = 0.68]{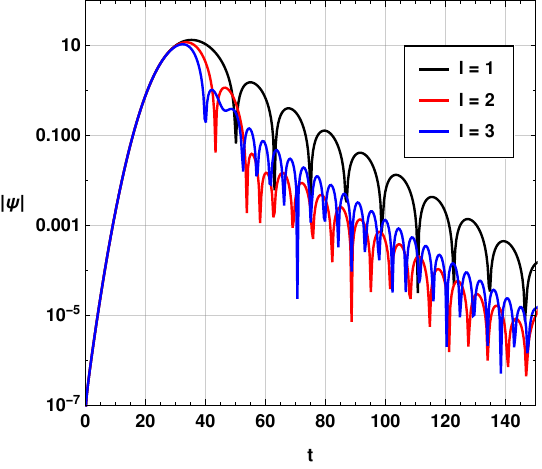}} 
   \vspace{-0.2cm}
\caption{The time-domain profiles of the massless scalar
perturbations (first panel) and electromagnetic perturbations (second panel) for different multipole moments $l$ with the parameter values $M=1$, $G = 1$, $n= 0$, $\alpha =0.3$, $\beta = 0.1$ and $\Lambda = 0.0$. }
\label{time04}
\end{figure}

\begin{figure}[htbp]
\centerline{
   \includegraphics[scale = 0.68]{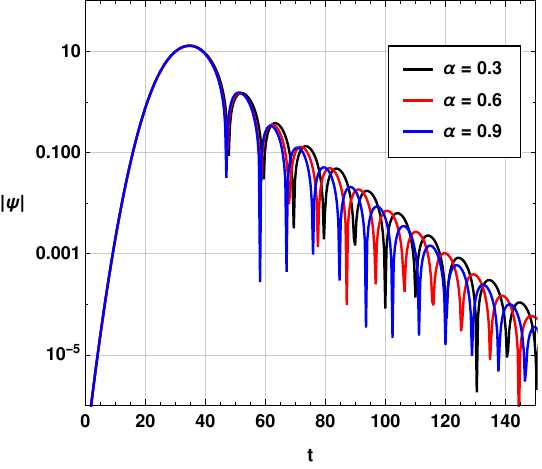}\hspace{0.5cm}
   \includegraphics[scale = 0.68]{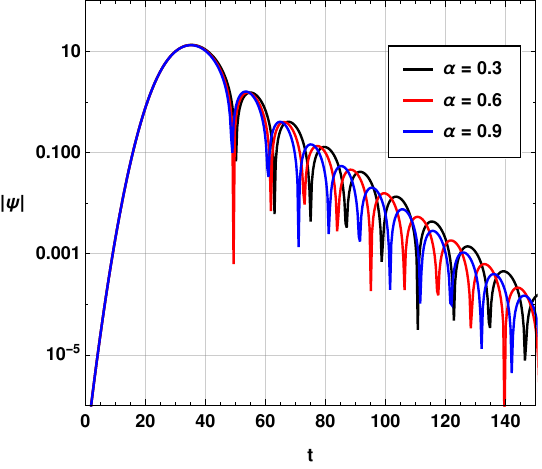}} 
   \vspace{-0.2cm}
\caption{The time-domain profiles of the massless scalar
perturbations (first panel) and electromagnetic perturbations (second panel) for different values of $\alpha$ with the parameter values $M=1$, $G = 1$, $n= 0$, $l =1$, $\beta = 0.1$ and $\Lambda = 0.0$. }
\label{time05}
\end{figure}

\begin{figure}[htbp]
\centerline{
   \includegraphics[scale = 0.68]{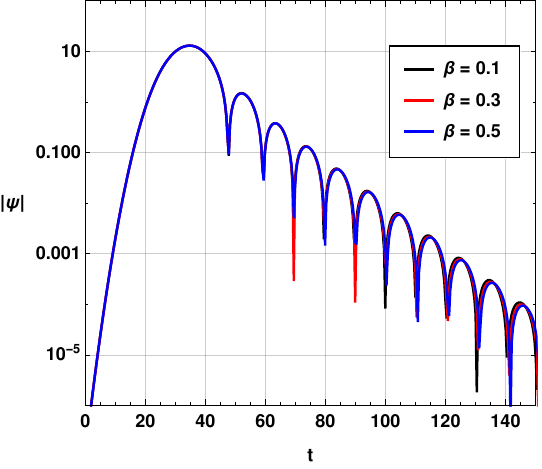}\hspace{0.5cm}
   \includegraphics[scale = 0.68]{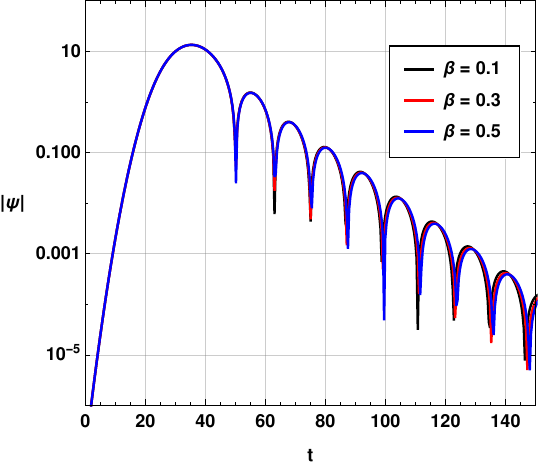}} 
   \vspace{-0.2cm}
\caption{The time-domain profiles of the massless scalar
perturbations (first panel) and electromagnetic perturbations (second panel) for different values of $\beta$ with the parameter values $M=1$, $G = 1$, $n= 0$, $\alpha =0.3$, $l = 1$ and $\Lambda = 0.0$. }
\label{time06}
\end{figure}

We have shown the time domain profiles for scalar and electromagnetic perturbations in Fig.s \ref{time01}, \ref{time02}, \ref{time03}, \ref{time04}, \ref{time05} and \ref{time06}. The time-domain profiles presented in these figures exhibit strong consistency with the QNM frequencies computed using the sixth-order WKB-Padé approximation, as listed in Table \ref{tab02}. For fixed deformation parameters $\alpha = 0.3$, $\beta = 0.2$, and $\Lambda = 0$, the scalar and electromagnetic perturbations display oscillation frequencies and damping rates in the time-domain evolution that closely match the real and imaginary parts of the fundamental ($n = 0$) QNMs. As the multipole index $l$ increases, the time-domain signals show higher oscillation frequencies and slower damping, which aligns well with the increasing real part and decreasing imaginary part of the QNM spectrum. Scalar perturbations consistently exhibit slightly higher frequencies and faster decay than electromagnetic ones, a trend that is evident in both the numerical evolution and the WKB results. Additionally, the influence of the deformation parameter $\alpha$ and the control parameter $\beta$ is reflected in the time-domain profiles: increasing $\alpha$ leads to more pronounced peaks and slightly delayed decay, while increasing $\beta$ tends to reduce the oscillation frequency and prolong the signal, indicating a flatter effective potential. These qualitative behaviors are in agreement with theoretical expectations from perturbation theory, even though the WKB data is only presented for fixed parameter values. Overall, the time-domain analysis not only corroborates the WKB-based QNM frequencies but also provides deeper insight into the dynamical response of deformed Schwarzschild black holes, reinforcing the reliability of both methods in capturing the essential features of scalar and electromagnetic field perturbations.}

\section{Greybody Factors}\label{Sec6}

In this section, we examine the calculation of greybody factors in the context of massless scalar and electromagnetic perturbations in black hole spacetimes. Building on Hawking's groundbreaking discovery in 1975, which revealed that black holes emit radiation known as Hawking radiation, subsequent research has demonstrated that this radiation is not completely black but rather has a greybody appearance due to modifications in the spectrum as it travels away from the black hole's horizon \cite{Hawking:1975vcx, Singleton:2011vh, Akhmedova:2008dz}. This phenomenon, characterized by the greybody factor, accounts for the redshift and other effects that alter the observed radiation compared to its original form near the event horizon.

To calculate these greybody factors, various methods have been developed and utilized in the literature, such as those by Maldacena et al. (1996) and Fernando (2004), among others \cite{Maldacena:1996ix, Fernando:2004ay, Okyay2022, Ovgun188, Pantig:2022gih, Yang:2022xxh, Panotopoulos:2018pvu, Panotopoulos:2016wuu, Rincon:2018ktz, Ahmed:2016lou, Javed:2022kzf, Al-Badawi:2022aby}. In this study, we employ the higher-order WKB (Wentzel-Kramers-Brillouin) approximation method to determine greybody factors for scalar and electromagnetic perturbations.

We begin by analyzing the wave equation \eqref{radial_scalar} under boundary conditions that allow for incoming waves from infinity. Given the symmetry of scattering properties, this is equivalent to considering the scattering of a wave originating from the black hole horizon. This setup is particularly useful for determining the proportion of waves reflected back towards the horizon by the potential barrier. The boundary conditions for this scenario are expressed as:
\begin{equation}\label{BC}
\begin{array}{ccll}
    \Psi &=& e^{-i\omega r_*} + R e^{i\omega r_*},& r_* \rightarrow +\infty, \\
    \Psi &=& T e^{-i\omega r_*},& r_* \rightarrow -\infty, \\
\end{array}%
\end{equation}
where $R$ and $T$ denote the reflection and transmission coefficients, respectively, and satisfy the conservation of probability:
\begin{equation}\label{1}
\left|T\right|^2 + \left|R\right|^2 = 1.
\end{equation}
The transmission coefficient $\left|T\right|$, which is equivalent to the greybody factor $A$ of the black hole, is calculated using the WKB approach:
\begin{equation}
\left|A\right|^2=1-\left|R\right|^2=\left|T\right|^2.
\end{equation}

In the WKB method, the reflection coefficient $R$ is given by:
\begin{equation}\label{moderate-omega-wkb}
R = (1 + e^{- 2 i \pi K})^{-\frac{1}{2}},
\end{equation}
where the phase factor $K$ is determined by the equation:
\begin{equation}
K - i \frac{(\omega^2 - \mathcal{V}_{0})}{\sqrt{-2 \mathcal{V}_{0}^{\prime \prime}}} - \sum_{i=2}^{i=6} \Lambda_{i}(K) =0.
\end{equation}
Here, $\mathcal{V}_0$ represents the maximum value of the effective potential, $\mathcal{V}_{0}^{\prime \prime}$ is its second derivative at the maximum, and $\Lambda_i$ denotes higher-order WKB corrections that depend on the potential's derivatives up to the $2i$th order at the maximum. These corrections, as discussed in works such as \cite{Schutz,Will_wkb,Konoplya_wkb,Maty_wkb}, are crucial for improving the accuracy of the WKB approximation. Although the 6th order WKB formula is the primary method used in this study, lower orders may be applied for lower frequencies and multipole numbers.

It is important to note that the WKB approach's accuracy diminishes at low frequencies, where the reflection is nearly total and the greybody factors approach zero. However, this limitation does not significantly affect the estimation of energy emission rates, which remains accurate under the WKB framework. Since the WKB method is a well-established technique, further details are omitted here; for more information, readers are referred to comprehensive reviews such as \cite{Konoplya:2019hlu, Konoplya:2011qq}.

\begin{figure}[htbp]
\centerline{
   \includegraphics[scale = 0.8]{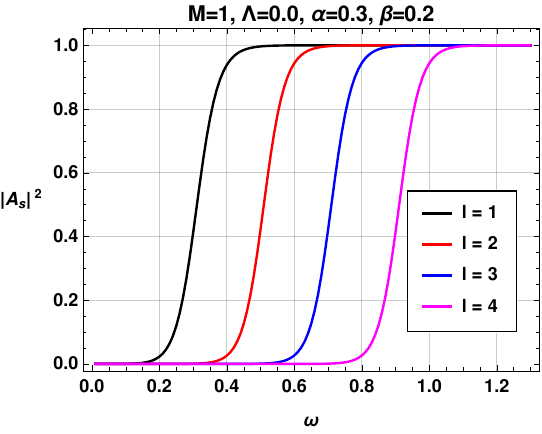}\hspace{0.5cm}
   \includegraphics[scale = 0.8]{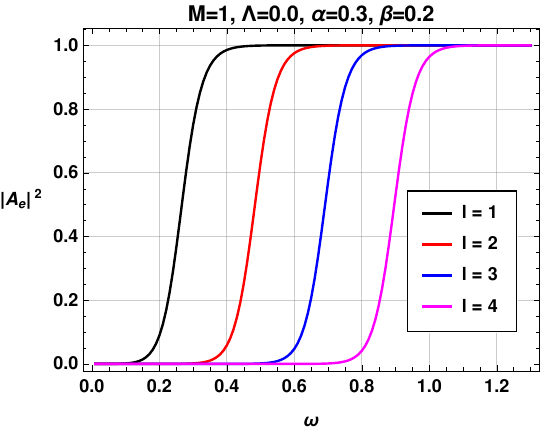}} \vspace{-0.2cm}
   
\caption{The greybody for scalar and electromagnetic perturbation with $\omega$}
\label{G01}
\end{figure}

\begin{figure}[htbp]
\centerline{
   \includegraphics[scale = 0.8]{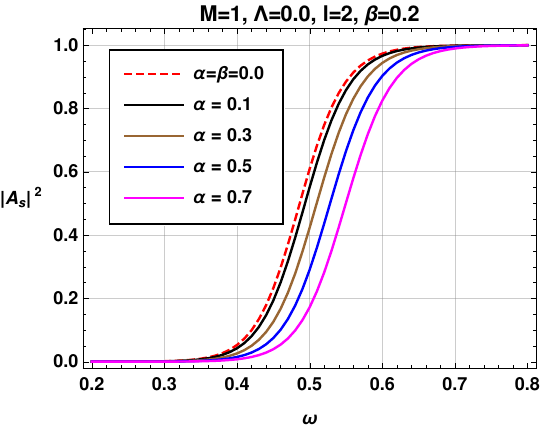}\hspace{0.5cm}
   \includegraphics[scale = 0.8]{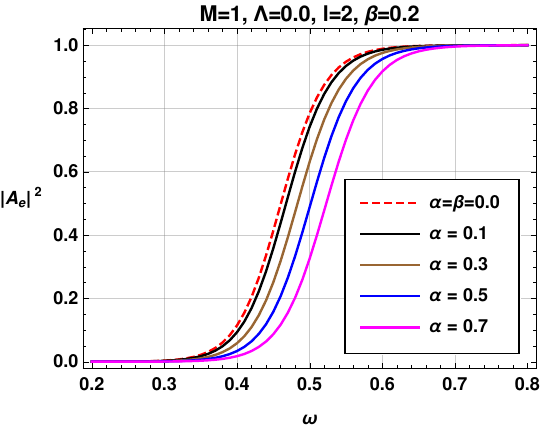}} \vspace{-0.2cm}
   
\caption{The variation of greybody for scalar and electromagnetic perturbation with $\omega$}
\label{G02}
\end{figure}

\begin{figure}[htbp]
\centerline{
   \includegraphics[scale = 0.8]{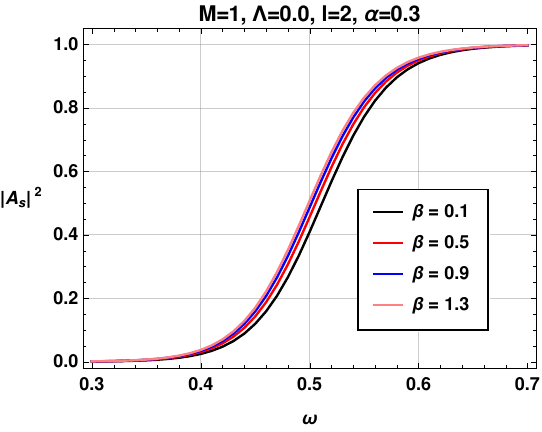}\hspace{0.5cm}
   \includegraphics[scale = 0.8]{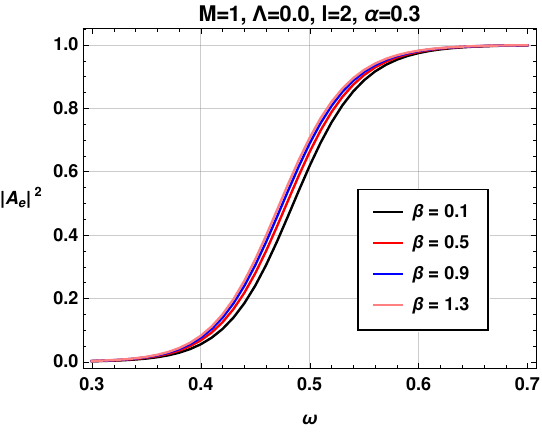}} \vspace{-0.2cm}
   
\caption{The variation of greybody for scalar and electromagnetic perturbation with $\omega$}
\label{G03}
\end{figure}

We have shown the variations of greybody factors with respect to the model parameters in Fig.s \ref{G01}, \ref{G02} and \ref{G03}.

These figures illustrate the behaviour of greybody factors for scalar and electromagnetic (EM) perturbations in deformed black holes. The plots explore the effects of various parameters, including the deformation parameter $\alpha$, control parameter $\beta$, and the angular momentum quantum number $ l $, on the absorption probability $ |A|^2 $ as a function of frequency $ \omega $. These figures shed light on how these parameters influence the transmission of perturbations through the potential barrier surrounding the black hole, offering insights into the black hole's interaction with its surroundings.

In Fig. \ref{G01}, the greybody factors are examined for different values of $ l $ and $\alpha$, with $\beta$ held constant. The two panels depict the behaviour of scalar and electromagnetic perturbations for $ l = 1 $ to $ l = 4 $, showing that higher angular momentum perturbations face a higher potential barrier, leading to a steeper ascent in the greybody factor curves at higher frequencies. This suggests that perturbations with greater angular momentum are more strongly scattered by the potential, reducing their likelihood of transmission at lower frequencies.

The two panels of Fig. \ref{G02} focus on the effect of varying $\alpha$, with $ l = 2 $ and $\beta = 0.2$ held constant. As $\alpha$ increases, the onset of significant transmission shifts to higher frequencies, indicating that the deformation of the black hole spacetime increases the potential barrier. This behaviour highlights the strong influence of the deformation parameter on the greybody factors, suggesting that modifications to black hole geometry can have a noticeable effect on the transmission characteristics of perturbations. { To understand the variation of the greybody factors from a standard black hole in the absence of deformation, we have shown a curve with $\alpha=\beta=0.$ It is clearly evident that the presence of deformation increases the potential barrier. }

Fig. \ref{G03} further explores the impact of the control parameter $\beta$, with fixed values for $\alpha = 0.3$ and $ l = 2 $. The greybody factors for scalar and electromagnetic perturbations are plotted for $\beta$ values ranging from 0.1 to 1.3. Here, the influence of $\beta$ is more subtle compared to $\alpha$ or $ l $, as the curves for different $\beta$ values are closely aligned. While increasing $\beta$ leads to a slightly earlier onset of transmission and a marginally sharper rise in $ |A|^2 $, the overall effect of $\beta$ on the greybody factor is minor in this frequency range.

When combining the results from all these three figures, it is clear that the angular momentum quantum number $ l $ and the deformation parameter $\alpha$ play a more prominent role in shaping the greybody factors, significantly affecting the absorption probabilities of perturbations. The control parameter $\beta$, while still influential, fine-tunes the greybody factor profiles without causing dramatic shifts. This suggests that the most notable observational signatures of deformed black holes might arise from variations in $ l $ and $\alpha$, while changes in $\beta$ would have more nuanced effects.

Overall, these results provide a comprehensive view of how greybody factors depend on key parameters in deformed black hole spacetimes. The findings have important implications for understanding the observational signatures of black holes and the influence of modified gravity theories on black hole properties. Further investigations into the interplay of these parameters could help refine models of black hole evaporation, quasinormal modes, and related phenomena, contributing to a deeper understanding of gravity in extreme regimes.

\section{Emission Rate}\label{Sec7}
The particle emission rate from a black hole is a critical optical property that provides insight into the black hole's stability and lifetime. This emission rate is intrinsically linked to two fundamental characteristics: the Hawking temperature and the black hole shadow. The Hawking temperature, which arises from quantum effects near the event horizon, governs the rate at which particles are thermally emitted by the black hole. Meanwhile, the shadow - the dark region against the backdrop of luminous sources - provides a geometric measure that influences the observational properties of black hole emissions. Together, these attributes enable a comprehensive understanding of the energy emission profile, which can be expressed as \cite{Wei2013, Cai:2020igv, Decanini:2011xi} 

\begin{eqnarray}
\frac{\partial ^2E}{\partial t\partial \omega }=\frac{\left(2 \pi ^2 \omega ^3\right) \sigma}{e^{\frac{\omega }{T_H}}-1}
\end{eqnarray}
where $\sigma=\pi r_{sh}^2$ represents the cross section, $r_{sh}$ and $T_H$ are obtained from Eqs. \eqref{rsh} and \eqref{TH}, respectively. Analyzing this emission rate allows researchers to evaluate the black hole’s dynamic properties and to predict its longevity, with significant implications for studying black hole thermodynamics and quantum gravity.
\begin{figure}[ht!]
\centerline{
   \includegraphics[scale = 0.48]{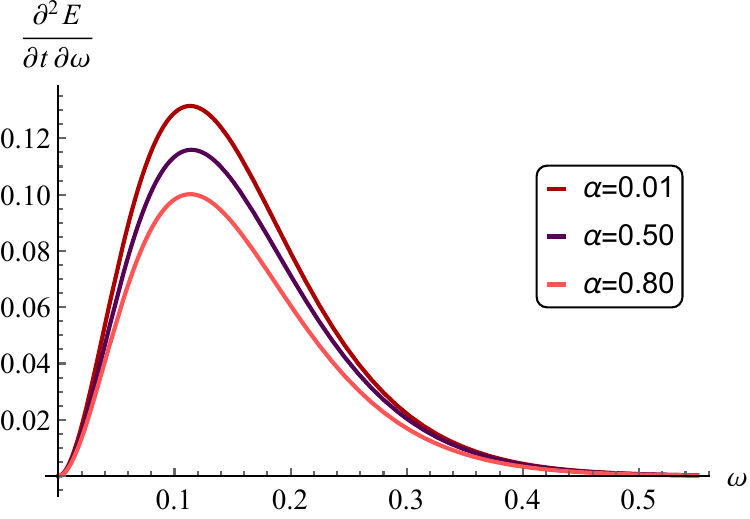}\hspace{0.5cm}
   \includegraphics[scale = 0.48]{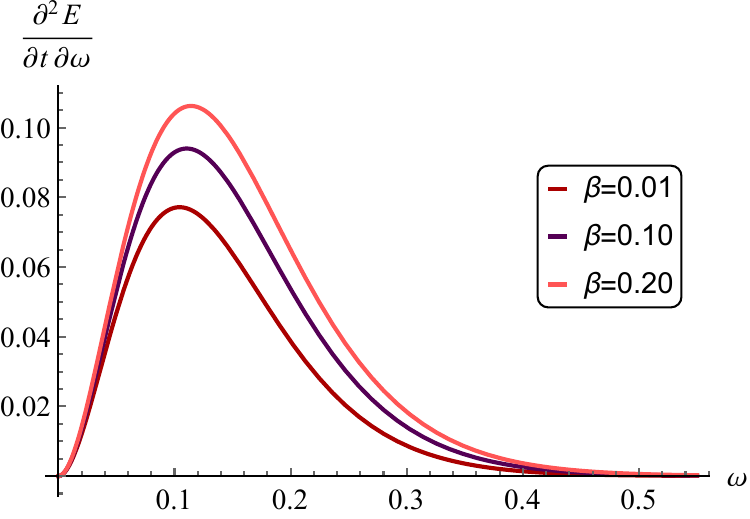}} \vspace{-0.2cm}
\caption{Variation of emission rate. Left panel: Parameters are set as $M=1,\,\beta=0.2,$ and $\Lambda=-0.002$. Right panel: Parameters are set as $M=1,\,\alpha=0.7,$ and $\Lambda=-0.002$}
\label{fig:EmissionRate}
\end{figure}
{ As shown in Fig. \ref{fig:EmissionRate}, an increase in the parameter $\alpha$, causes a decrease in the emission rate of the black hole. As a result, the lifetime of the black hole increases in comparison to that of a black hole without deformation.
On the other hand, an increase in the parameter $\beta$, leads to an increase in the emission rate and as a result, the life span of the black hole decreases.  }

\section{Topological Characteristics}\label{Sec9}
{
In this section, the stability of the photon sphere and the classification of black holes based on thermodynamic potentials will be investigated using a topological approach.
}
\subsection{Topological Charge of Photon Sphere}
\begin{figure}[htbp!]
\centerline{
   \includegraphics[scale = 0.48]{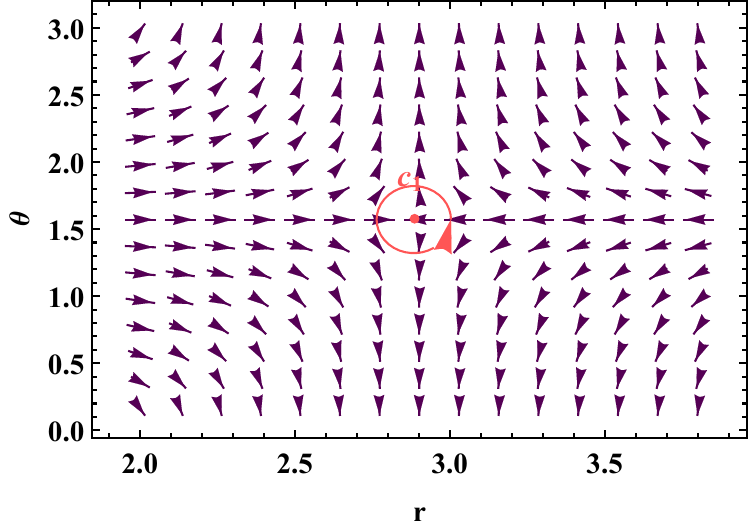}\hspace{0.5cm}
   \includegraphics[scale = 0.5]{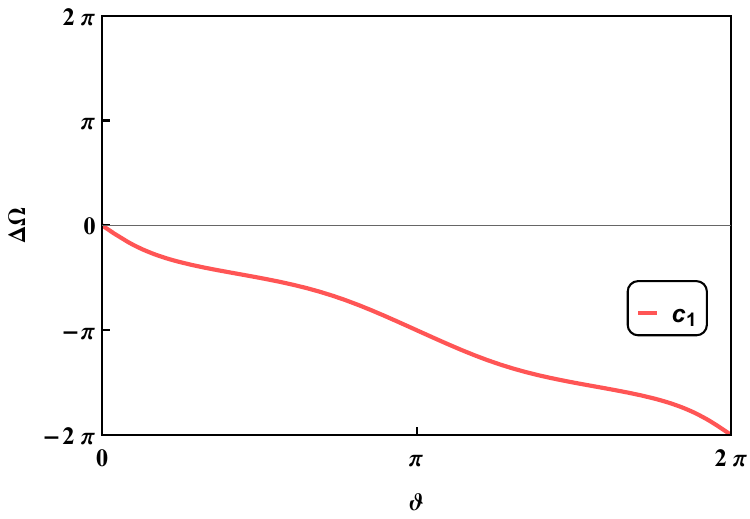}} \vspace{-0.2cm}
\caption{Assuming $M=1$, $\alpha=0.20$, $\beta=0.20$, and $\Lambda=-0.002$. Left panel: Vector space of potential $H(r,\,\theta)$, zero point is located at $r_c=2.88232$. Right panel: Variation of $\Delta\Omega$ versus $\vartheta$, considering $a=b=0.3$.}
\label{fig:TopoPS}
\end{figure}

{Photon sphere of the black hole of Eq. \eqref{final_line_element} is explored in Sec. \ref{Sec4}. In order to study the stability or instability of the photon sphere, a topological approach is employed. To this end, a potential of form \cite{wei2020topological,sadeghi2024role}
\begin{equation}
H(r,\,\theta)=\frac{1}{\sin\theta}\frac{\mathcal{F}(r)}{r^2}
\end{equation}
is defined, which can be represented in a vector space using \cite{wei2020topological}
\begin{eqnarray}
\phi^H=\left(\frac{\partial}{\partial r} H(r,\,\theta),
\,\frac{\partial}{\partial \theta} H(r,\,\theta)\right)
\end{eqnarray}
which can be normalized as
\begin{eqnarray}
n^H=\frac{1}{||\phi^H||}\,(\phi^H_r,\,\phi^H_\theta).
\end{eqnarray}
Each photon sphere in vector space is represented as a zero point, to which a topological charge can be assigned. To study the stability of the zero point, the sign of the topological charge arising from the rotation of vectors around this point is examined as \cite{zhu2024topological}
\begin{eqnarray}
w_i=\frac{1}{2\pi}\oint_c d(\arctan\frac{n_\theta}{n_r}).
\end{eqnarray}
For this purpose, a closed contour is taken around the zero point $r_i$, and a change of variables in the form of
\begin{eqnarray}
r=a \cos\vartheta +r_i,\;\;\;\;
\theta=b \sin\vartheta+\frac{\pi}{2},
\end{eqnarray}
is utilized to calculate the topological charge of the zero point. Fig. \ref{fig:TopoPS} illustrates the vector space of potential $H(r,\,\theta)$ for the case $M=1$, $\alpha=0.20$, $\beta=0.20$, and $\Lambda=-0.002$, and as shown, a zero point is located at $r_c=2.88232$.
}
{Considering $a=b=0.3$, the topological charge of the photon sphere located at $r_c$ equals -1, indicating  an unstable photon sphere. This instability is consistent with our studies in Sec. \ref{Sec4}, as this unstable photon sphere leads to the formation of the shadow.}
\subsection{Topological Charge of Thermodynamic Potentials}
\begin{figure}[ht!]
\centerline{
   \includegraphics[scale = 0.4]{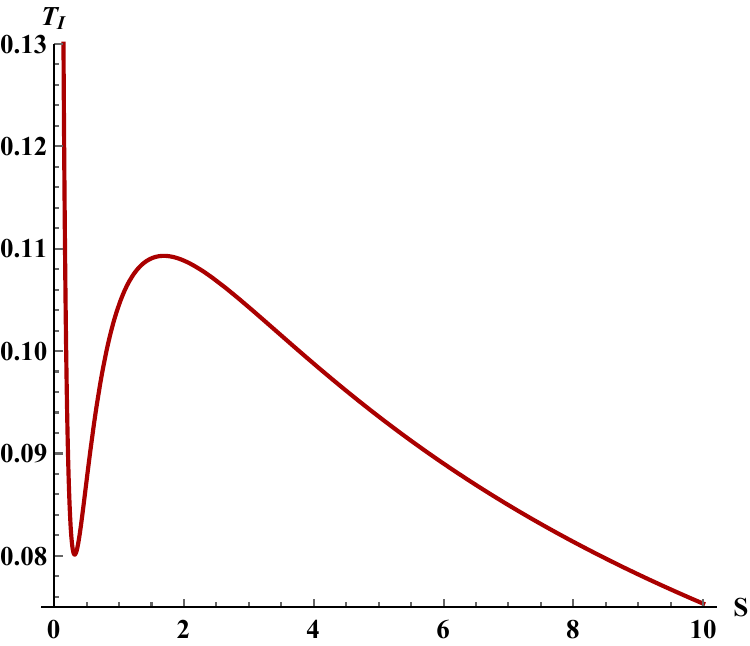}\hspace{0.1cm}
   \includegraphics[scale = 0.48]{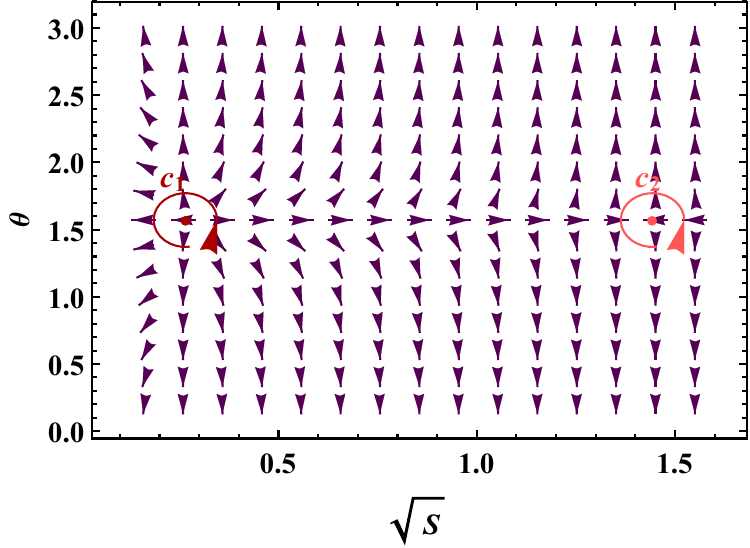}\hspace{0.1cm}
   \includegraphics[scale = 0.5]{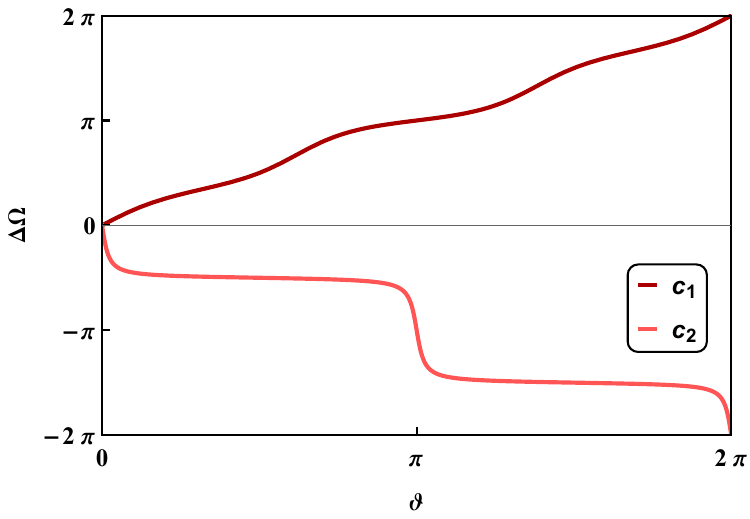}} \vspace{-0.2cm}
\caption{Assuming $\alpha=0.40$ and $\beta=0.18$. Left panel: Isobaric curve of Hawking temperature. Middle Panel: Vector space of potential $\Phi$. Right panel: Variation of $\Delta\Omega$ versus $\vartheta$, considering $a=b=0.1$.}
\label{fig:TopoTemp}
\end{figure}

{Similar to the previous section, the thermodynamic potentials of the black hole can be studied using a topological approach. Hawking temperature of the black hole was calculated in Eq. \eqref{TH}. This temperature can be expressed in isobaric form as
\begin{eqnarray}
T_i=\frac{1}{2 \sqrt{\pi S}}-\frac{\sqrt{\pi } \alpha  S}{\left(\sqrt{\pi } \beta +\sqrt{S}\right)^5}
\end{eqnarray}
The curve of isobaric temperature variations as a function of $\sqrt{S}$ is presented in Fig. \ref{fig:TopoTemp} for the case $\alpha=0.40$ and $\beta=0.18$.}
{The Hawking temperature--dependent potential is defined as \cite{wei2022topology,mehmood2023thermodynamic}
\begin{eqnarray}
\Phi =\frac{1}{\sin \theta} T_{i}.
\end{eqnarray}
The vectors of the above potential can be represented using \cite{wei2022topology}
\begin{equation}
\phi^{\Phi}
=(\frac{\partial}{\partial r_h}\Phi,\,\frac{\partial}{\partial \theta}\Phi),
\end{equation}
This vector space for cases $\alpha=0.40$ and $\beta=0.18$ is depicted in Fig. \ref{fig:TopoTemp}. This vector space has two zero points enclosed by contours $c_1$ and $c_2$. Considering $a=b=0.1$, their topological charges are +1 and $-1$, respectively, indicating a novel phase transition and a conventional phase transition \cite{wei2022topology,yerra2022topology}.}

\begin{figure}[ht!]
\centerline{
   \includegraphics[scale = 0.46]{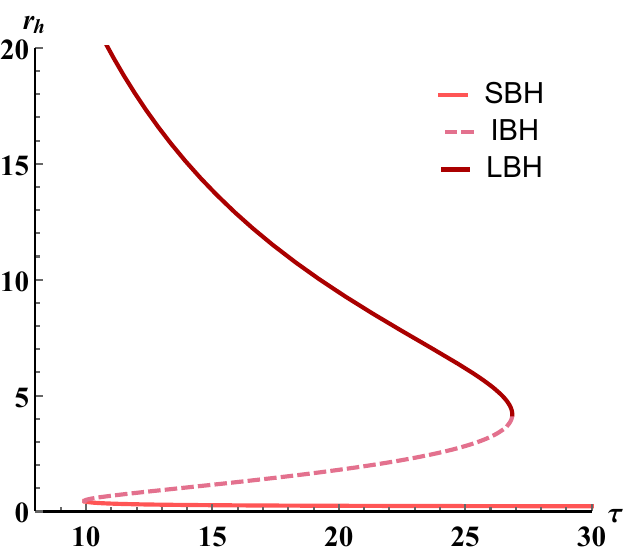}\hspace{0.1cm}
   \includegraphics[scale = 0.48]{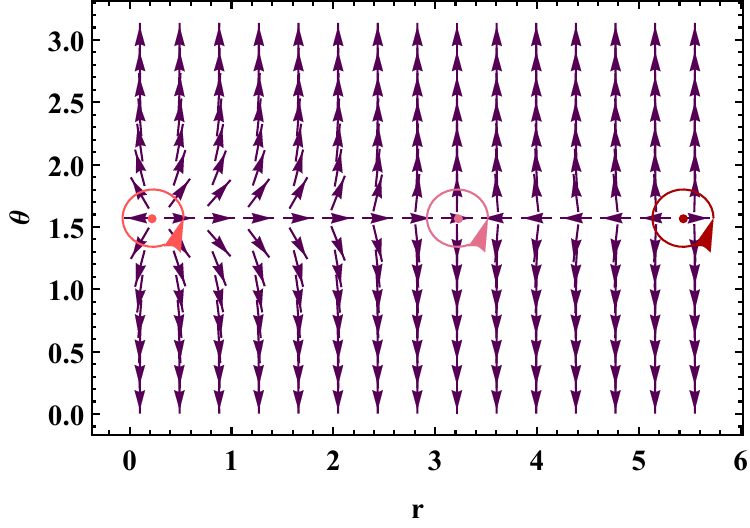}\hspace{0.1cm}
   \includegraphics[scale = 0.5]{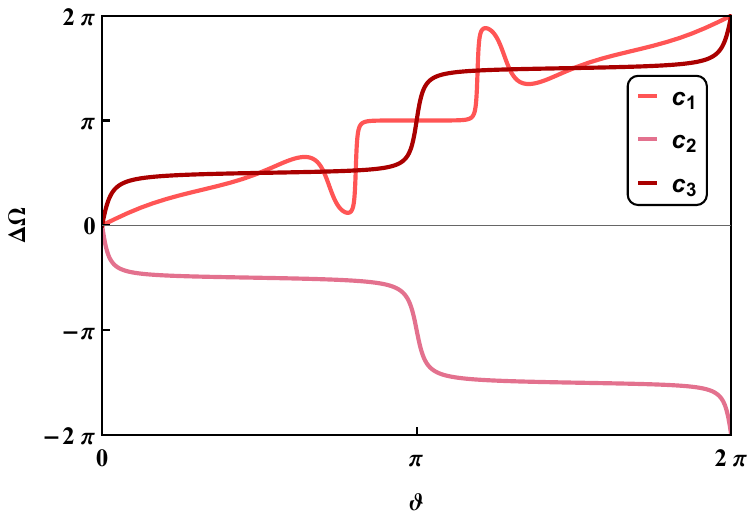}} \vspace{-0.2cm}
\caption{Assuming $\alpha=0.20$, $\beta=0.20$, and $P=0.0022$. Middle panel: Vector space of potential $\mathcal{F}$, when $\tau=26$. Right panel: Variation of $\Delta\Omega$ versus $\vartheta$, considering $a=b=0.3$.}
\label{fig:TopoFree}
\end{figure}

{Generalized free energy of a black hole is another quantity which can be analyzed thermodynamically and is defined as \cite{wei2022black,wu2025novel}
\begin{eqnarray}
F=M(r_h) - \frac{S}{\tau}
\end{eqnarray}
that $M(r_h)$ is black hole's mass as a function of horizon radius, $\tau$ is the inverse of black hole temperature, and on the horizon becomes $\tau\big|_{r=r_h}=T_H$, and $S$ indicates entropy of the black hole and is calculated as
\begin{eqnarray}
S=\int \frac{d M(r_h)}{T} d r_h = \pi r_h^2
\end{eqnarray}
Thus, off--shell generalized free energy of the black hole of form Eq. \eqref{final_line_element} reads
\begin{equation}
F=\frac{r_h}{2}+\frac{\alpha  \left(\beta ^2+3 r_h^2+3 \beta  r_h\right)}{(\beta +r_h)^3}-\frac{\Lambda  r_h^3}{6}-\frac{\pi  r_h^2}{\tau }
\end{equation}
and vectors of the above potential are defined as \cite{wei2022black,fang2023revisiting}
\begin{equation}
\phi^{F}=(\frac{\partial}{\partial r_h}F,\,-\cot \theta \csc \theta).
\end{equation}
Zero point of potential field $F$ is located at $\dfrac{\partial}{\partial r_h}F=0$. Thus parameter $\tau$ at the zero point can be found as function of $r_h$ as
\begin{equation}
\tau=\frac{4 \pi  r_h (\beta +r_h)^4}{\beta ^4+r_h^4+4 \beta  r_h^3-\alpha  r_h^2+6 \beta ^2 r_h^2-\Lambda  r_h^2 (\beta +r_h)^4+4 \beta ^3 r_h}
\end{equation}
Fig. \ref{fig:TopoFree} illustrates $r_h-\tau$ curve for the given set of initial values $\alpha=0.20$, $\beta=0.20$, and $P=0.0022$. As shown, As can be seen, this curve changes direction at $(0.45782,\,9.94262)$ and $(4.18812,\,26.8591)$.}
{By choosing $\tau=26$, the vector space of the potential $F$ is illustrated in Fig. \ref{fig:TopoFree}. This vector space includes three zero points at $r_{h_1}=0.23014$, $r_{h_2}=3.21936$, and $r_{h_3}=5.43388$, with their topological charges equal to $+1$, $-1$, and $+1$, respectively, where $a=b=0.3$. For the given set of initial values, the whole topological charge is $+1$. Therefore, this black hole is classified within a class analogous to RN--AdS black hole \cite{wei2022black}.
}
\section{Concluding Remarks}\label{Sec8}
In this work, we explored several properties of the deformed black hole spacetime, such as the temperature, photon and shadow radii, QNMs, greybody factors and emission rate, mainly focusing on the effects of the deformation parameter $ \alpha $ and control parameter $ \beta $ on both scalar and electromagnetic perturbations.
The investigation of shadow radius as a function of parameters $\alpha,\,\beta$ and $\Lambda$, leads to an upper and lower limit of parameters in which the shadow radius is allowed as a result of the observed data of Sgr A$^*$.
Our analysis of QNMs revealed that an increase in $ \alpha $ significantly enhances the oscillation frequency and damping rate of ringdown GWs while increasing $ \beta $ leads to a non-linear behaviour in both frequency and damping rates. Specifically, the damping rate initially increases with $ \beta $ before reversing its trend beyond $ \beta = 0.65 $. For electromagnetic perturbations, we observed a similar pattern, though with slightly lower oscillation frequencies and damping rates compared to scalar perturbations. These variations underscore the sensitivity of black hole dynamics to these parameters, which are crucial for understanding astrophysical processes like black hole mergers and gravitational wave emissions. 
{ This investigation ascertains that deformation parameters $\alpha$ and $\beta$ play distinct roles in shaping the black hole geometry and, consequently, influence the behaviour of QNM in different ways. The parameter $\alpha$ governs the overall strength of the deformation introduced through the energy density profile, effectively modifying the depth and curvature of the effective potential experienced by perturbations. As $\alpha$ increases, the potential barrier becomes steeper and more localized, leading to higher oscillation frequencies ($\omega_R$) and faster damping rates ($|\omega_I|$), as is typical in systems with sharper confinement. This is consistent with the intuitive expectation that stronger spacetime curvature near the peak of the potential enhances the restoring force on perturbations, thus increasing the QNM frequency.

On the other hand, $\beta$ controls the regularization scale near the black hole core and has a more subtle influence. Its primary effect is to alter the inner structure of the geometry without significantly modifying the asymptotic form. As shown in the results, the behaviour of $\omega_I$ with $\beta$ is non-monotonic, displaying a reversal near a critical value (e.g., $\beta \approx 0.65$), which can be interpreted as a transition in the internal structure of the effective potential. For small $\beta$, the regularizing effect is weak, and the potential resembles that of a singular black hole; increasing $\beta$ initially strengthens the damping, but beyond a threshold, the regularization dominates and leads to a broadening of the potential barrier, reducing the decay rate. This non-linear behaviour reflects the competition between enhanced regularity and the geometric sharpness of the potential barrier, and it may have interesting implications for the stability and ringdown profiles of regular or exotic compact objects.}

{ While our study reveals that the deformation parameters $\alpha$ and $\beta$ introduce significant modifications to the standard QNM spectrum -particularly enhancing the oscillation frequency and altering the damping behavior—translating these theoretical results into observational constraints presents a challenge with current detection capabilities. Ground-based gravitational wave detectors such as LIGO and Virgo are primarily sensitive to QNM frequencies within the range $f \in [12 \, \text{Hz}, 1.2 \, \text{kHz}]$. By converting the dimensionless QNM frequencies into physical units using the relations derived in Ref. \cite{Ferrari2008}, we estimate the mass ranges of black holes whose ringdown signals could be detected within this sensitivity window. Our analysis shows that, for both types of perturbations (scalar, and electromagnetic), stellar-mass black holes with masses between approximately $8 M_{\odot}$ and $1100 M_{\odot}$ are within the detectable range for LIGO-Virgo. However, for supermassive black holes such as Sagittarius A\* ($M \sim 3.7 \times 10^6 M_{\odot}$), the corresponding QNM frequencies fall well below the LIGO-Virgo band but lie comfortably within the future LISA frequency range ($0.1 \, \text{mHz}$ to $1 \, \text{Hz}$). Therefore, although current ground-based detectors may not provide stringent constraints on the deformation parameters, space-based observatories like LISA offer promising avenues for probing such modifications through the precise measurement of QNMs from massive black holes. In this context, our results provide a theoretical basis for identifying potential signatures of deviations from general relativity, such as those induced by deformation parameters, in upcoming gravitational wave observations.}

Additionally, we examined the greybody factors for scalar and electromagnetic perturbations, with a focus on the role of angular momentum $ l $, $ \alpha $, and $ \beta $. Our results indicate that both $ l $ and $ \alpha $ play a dominant role in shaping the transmission of perturbations through the black hole’s potential barrier. Higher angular momentum leads to stronger scattering, while increasing $ \alpha $ shifts the onset of significant transmission to higher frequencies. In contrast, $ \beta $ has a more subtle impact, fine-tuning the transmission profiles without causing major shifts in the greybody factors.

{ We have also investigated the topological characteristics of the black hole. We found that this black hole can be classified within a class analogous to RN--AdS black hole from this perspective.}

Overall, our findings highlight the intricate interplay between deformation parameters and the behaviour of black holes under perturbations. The results provide valuable insights into how modified black hole geometries affect QNMs, damping rates, and greybody factors, with implications for both theoretical studies of black holes and their observational signatures in gravitational wave astronomy. Future work could focus on exploring these effects within different modified gravity frameworks to further our understanding of black hole evolution and energy dissipation in extreme regimes.

\section*{Acknowledgments}
The authors would like to thank the referees for their constructive comments. DJG acknowledges the contribution of the COST Action CA21136  -- ``Addressing observational tensions in cosmology with systematics and fundamental physics (CosmoVerse)". F. S. and
H. H. are grateful to the Excellence project FoS UHK 2203/2025-2026 for the financial
support. Also, the research of H.H. was supported by the Q-CAYLE project, funded by the European Union-Next Generation UE/MICIU/Plan de Recuperacion, Transformacion y Resiliencia/Junta de Castilla y Leon (PRTRC17.11), and also by project PID2023-148409NB-I00, funded by MICIU/AEI/10.13039/501100011033. Financial support of the Department of Education of the Junta de Castilla y Leon and FEDER Funds is also gratefully acknowledged (Reference: CLU-2023-1-05). The authors would like to thank Prof. R. A. Konoplya for his useful discussion.

\section*{Data Availability Statement}
There are no new data associated with this article.

\end{document}